\documentclass[preprint,review,12pt]{elsarticle}

\usepackage{amssymb}

\usepackage{amsmath}
\usepackage{cases}
\usepackage{mathtools}
\usepackage[ruled,vlined]{algorithm2e}
\usepackage[capitalise]{cleveref} 

\usepackage{lineno}

\newcommand{\bT}{\ensuremath{\mathbf{T}}\xspace}
\newcommand{\bE}{\ensuremath{\mathbf{E}}\xspace}
\newcommand{\bQ}{\ensuremath{\mathbf{Q}}\xspace}
\newcommand{\bA}{\ensuremath{\mathbf{A}}\xspace}

\newcommand{\bB}{\ensuremath{\mathbf{B}}\xspace}
\newcommand{\bD}{\ensuremath{\mathbf{D}}\xspace}

\newcommand{\Tyy}{\ensuremath{T_{yy}}\xspace}
\newcommand{\Txx}{\ensuremath{T_{xx}}\xspace}
\newcommand{\Txy}{\ensuremath{T_{xy}}\xspace}

\newcommand{\bq}{\ensuremath{\mathbf{q}}\xspace}
\newcommand{\bx}{\ensuremath{\mathbf{x}}\xspace}
\newcommand{\bn}{\ensuremath{\mathbf{n}}\xspace}
\newcommand{\bt}{\ensuremath{\mathbf{t}}\xspace}

\newcommand{\bd}{\ensuremath{\mathbf{d}}\xspace}
\newcommand{\bu}{\ensuremath{\mathbf{u}}\xspace}
\newcommand{\bv}{\ensuremath{\mathbf{v}}\xspace}

\newcommand{\dx}[1]{\frac{\partial #1}{\partial x}}
\newcommand{\dy}[1]{\frac{\partial #1}{\partial y}}

\newenvironment{gequation} 
    {\begin{equation}
    \begin{gathered}
    }
    { 
    \end{gathered} 
    \end{equation}
    }
    
\newenvironment{bmat} 
    {\begin{bmatrix}
    }
    { 
    \end{bmatrix}
    }

\journal{International Journal of Solids and Structures}

\begin{document}

\begin{frontmatter}



\title{On discontinuities when computing the stress-field from the strain: A finite volume discretization}

\affiliation[label1]{organization={Department of Civil and Environmental Engineering\\
  Massachusetts Institute of Technology},
}
\affiliation[label2]{organization={Department of Materials Science and Engineering\\
   Massachusetts Institute of Technology},
}
\author[label1]{Benjamin C. Cameron}
\author[label2]{Cem C. Tasan}
\ead{tasan@mit.edu}




\begin{abstract}
This article focuses on developing a finite volume discretization for a recently-developed and widely-applicable system of PDEs, enabling computation of the full-field stress from a measured strain field and traction boundary conditions. This is applicable to datasets obtained via DIC, and applies when the material properties are unknown. Interestingly, the strain or strain-rate discontinuities at the interfaces between finite volumes cause reflection and transmission of the principal stresses. Hence, we analytically analyze the solution structure in this case, enabling the development of the discretization. In addition, analysis is conducted translating knowledge of the traction at the boundary into boundary conditions, so that the numerical method can be applied to a variety of loading conditions. The finite volume discretization is validated using strain fields obtained from finite element simulations. The developed discretization enables the system of equations to be applied to irregular geometries, finite deformation, general boundary conditions, and avoids instabilities that arise due to the Courant-Friedrichs-Lewy condition, significantly increasing the range of applications for which these PDEs can be applied.
\end{abstract}


\begin{highlights}
\item Developed computational method enables determination of constitutive equations from digital image correlation (DIC).
\item Theoretical solution obtained for discontinuity giving principal stress reflection.
\item Applies to isotropic elastic, elastoplastic, plastic and viscous deformation.
\item Alternative procedure to the virtual fields method and finite element model updating.
\item Demonstrated computation of stress-strain relationship for strains on the order of 100\%.
\end{highlights}

\begin{keyword}
constitutive behavior  \sep characteristics \sep mechanical testing \sep numerical algorithms \sep inverse problem
\end{keyword}

\end{frontmatter}




\section{Introduction}
\label{sec:intro}
The availability and quality of full-field deformation data have substantially improved in recent years. Image analysis algorithms, such as digital image correlation (DIC), have become more accurate \citep{Pan2009,Hild2006}, and there has been a rapid increase in software and hardware availability \citep{Blaber2015}. These approaches can be used to quantify complex plastic deformations, such as shear band formation, necking, and high strain-rate deformation. Although extensive full-field strain data is often available, the full-field stress typically remains unknown. Obtaining the full-field stress would enable (i) quantification of material properties when heterogeneous plastic flow instabilities arise; (ii) computation of material properties as a function of position; and (iii) quantification of stress partitioning in complex microstructures and composite materials. This motivates solutions to the inverse problem: computing the full-field stress from the strain when the constitutive equations remain unknown. In contrast, the forward problem is defined as computing the stress and strain fields from known constitutive relations and boundary conditions.

Inverse problem approaches target solutions consistent with force balance, the observed deformation field, and an assumed constitutive equation form. However, these constraints typically overdetermine the problem, so various optimization procedures are developed to minimize error. For example, one can minimize the difference between the assumed and computed deformation field using finite element (FE) model updating \citep{Rethore2010,Siddiqui2017,Viala2018}, or error in the force-balance equation using the virtual fields method \citep{Crouzeix2009,Pierron2012}. Other approaches minimize the difference between two stress fields, one consistent with force equilibrium, and one consistent with the assumed constitutive equation form and observed strain field. These include the constitutive equation gap method \citep{Florentin2010,Florentin2011}, and the data-driven method recently introduced \citep{Leygue2018,Stainier2019,Langlois2022}. These optimization-based methods can arrive at accurate solutions in the case of homogeneous material properties, for example, \citet{Fletcher2021} compute the parameters of various isotropic elasto-plastic constitutive equations with accuracies ranging between 0.07 and 5.51~\%. However, accuracies are inconsistent and depend on the specific problem, solution implementation, and assumptions made, for example, \citet{Leygue2018} arrive at an average stress-field error of 8.5~\% and maximum error greater than 50~\% for a case of homogeneous isotropic non-linear elasticity. Furthermore, these methods are often computationally intensive (or unfeasible) due to the non-linearity of the problem and the large number of iterations required.  The problems of accuracy and computational expense are exacerbated in heterogeneous materials, where errors can be on the order of 10-50~\% (e.g. \cite{Florentin2010,Rahmani2014}). The reader can refer to \citet{Pierron2021} for an extensive review of these approaches.

An alternative deterministic approach has recently been developed by the authors \citep{Cameron2021}, and subsequently by \citet{Liu2021}, which allows one to compute the stress field by solving a system of linear hyperbolic partial differential equations. The key assumption used to derive these equations is that the principal directions (eigenvectors) of the stress are known at each point from the observed deformation. This is straightforward in the case of isotropic (potentially non-linear) elasticity, or isotropic plastic deformation when the elastic strain is negligible (see \cref{ssec:background}). Given this assumption, the Cauchy stress, \bT, will satisfy the following system of equations:
\begin{subequations}
\begin{gather}
    \mathrm{div}(\mathbf{T}) + \mathbf{b} = \rho\ddot{ \mathbf{x}},
    \label{eq:system0a}\\
    [\mathbf{Q^\top TQ}]_{ij} = 0 \qquad \mathrm{for} \quad i\neq j,
    \label{eq:system0b}
\end{gather}
\label{eq:system0}
\end{subequations}
where \bQ is the matrix that diagonalizes \bT, $\mathbf{b}$ is the body force, $\ddot{\mathbf{x}}$ is the acceleration, and $\rho$ is the density\footnote{Notation: We use bold uppercase letters to denote second order tensors ($\mathbf{A},\mathbf{B}$,...), bold lowercase letters to denote vectors ($\mathbf{a},\mathbf{b}$,...), unbolded letters to denote scalars ($a$, $b$, $...$),  and a superscript $\top$ to denote the transpose ($\mathbf{A}^\top$, $\mathbf{a}^\top $, $...$). We use standard matrix notation, for example, we have $\mathbf{a}\cdot\mathbf{b} = \mathbf{a}^\top \mathbf{b}$ and $\mathbf{a}\otimes\mathbf{b} = \mathbf{a}\mathbf{b}^\top $. An over bar denotes functional dependence, e.g. $\mathbf{A} = \bar{\mathbf{A}}(\mathbf{b})$. We denote the first and second material time derivative of a second order tensor $\mathbf{\dot{A}}$ and $\mathbf{\ddot{A}}$ respectively. $\mathrm{Div}(\mathbf{A})$ and $\mathrm{div}(\mathbf{A)}$ correspond to the divergence of $\mathbf{A}$ with respect to the initial and deformed reference frame respectively. $\mathrm{Grad}(\mathbf{A})$ and $\mathrm{grad}(\mathbf{A)}$ correspond to the gradient of $\mathbf{A}$ with respect to the initial and deformed reference frame respectively. $|\mathbf{A}|=\sqrt{A_{ij}A_{ij}}$ gives the magnitude of $\mathbf{A}$ (Einstein summation notation is used). We denote the deviatoric part of a second order tensor $\mathbf{A}$ as $\mathbf{A_0}$, and the trace as $\mathrm{tr}(\mathbf{A})$. The spectral decomposition of a symmetric tensor $\mathbf{A}$ is written as $\mathbf{A} = \mathbf{Q_\bA\Lambda_\bA Q_\bA^\top }$ where $\mathbf{Q_\bA}$ has columns with the eigenvectors of $\mathbf{A}$ denoted $\{\bq_\bA^{(i)}|i=1,2,3\}$ (principal directions), and $\mathbf{\Lambda_A}$ is a diagonal matrix containing the eigenvalues $\{\lambda_\bA^i|i=1,2,3\}$ (principal values). This decomposition is also expressed as  $\mathbf{A} = \sum_i\lambda_\bA^{(i)}\bq_\bA^{(i)}\bq_\bA^{(i)\top}$. As the spectral decomposition is used extensively for the stress we use special notation: $\bT = \sum_i\sigma^{(i)}\bq^{(i)}\bq^{(i)\top}$ where $\{\sigma^{(i)}|i=1,2,3\}$ are the principal stresses and $\{\bq^{(i)}|i=1,2,3\}$ are the principal directions.} \citep{Cameron2021}. These variables are generally a function of position and time, however, these arguments will often be suppressed to avoid clutter (e.g. $\bT = \bar{\bT}(\bx,t)$, $\bQ = \bar{\bQ}(\bx,t)$, etc.). \cref{eq:system0a} is force balance and \cref{eq:system0b} arises from the constraint on the principal directions of \bT. The basic workflow envisioned is (i) a deformation experiment is conducted, (ii) the full-field kinematics are extracted using a technique such as DIC, (iii) the principal directions of stress are extracted from the kinematic fields, (iv) the stress field is obtained by solving the equations, (v) useful information, such as material properties, are extracted in post-processing. Step (iii) generally requires isotropy assumptions to determine the principal directions of the stress and traction boundary conditions are required to solve the equations. As discussed in \citet{Cameron2021}, this system of equations can be applied to a wide range of materials and deformation regimes including isotropic elastic solids, granular materials, plastic solids consistent with co-directionality, a subset of other plastic solids, Newtonian fluids, and a subset of non-Newtonian fluids. Furthermore, these equations have been generalized to apply to isotropic elasto-plastic deformation when the elastic strain is non-negligible (\citeauthor{CameronUnpublished2}, submitted). One limitation is that these equations cannot be directly applied to deformations with repeated stress eigenvalues. Addressing this will likely be the subject of future work.

Before this approach can be widely applied, a robust numerical procedure must be developed to solve the system of equations. While FE discretizations are well suited to parabolic or elliptic systems of partial differential equations, such as static linear elasticity, they are not well suited to hyperbolic systems, such as \cref{eq:system0}. Hyperbolic systems describe wave-like or transport phenomena, specifically, information propagates from the boundary along characteristic curves throughout the domain. The variable(s) at each point are influenced only by the variable(s) in the domain of dependence, which is controlled by the characteristic curves and the direction of information propagation. This imposes restrictions on the discretization, such as the Courant–Friedrichs–Lewy condition (\cite[chap. 6]{Strang2007}), and often gives rise to numerical instabilities if not properly accounted for. 

Finite difference discretizations are commonly used for hyperbolic systems of equations with a rectangular computational mesh (e.g. \cite{Alford1974}). Indeed, in \citet{Cameron2021}, a finite difference discretization was successfully applied to solve \cref{eq:system0} for a rectangular domain subject to infinitesimal deformation. However, difficulties arise when applying finite difference discretizations to solve \cref{eq:system0} with an irregular computational mesh or domain \citep{Strang2007}. This is not just problematic for computing the stress field in irregularly shaped objects, but also rectangular domains subject to finite heterogeneous deformation. Additional problems arise because the direction of the characteristic curves with respect to the chosen coordinates may change throughout the domain. This is similar to the steady state advection equation (e.g. \cite{Molenkamp1968}), but is in contrast to other systems such as elastic wave propagation (e.g. \citet{Karal1959}). This means that the implicit and explicit time stepping techniques, used for the latter case to resolve instabilities, cannot generally be applied to \cref{eq:system0}. The finite difference discretization applied in \citet{Cameron2021} relied on the characteristic curve orientations that were specific to that deformation, and becomes unstable when applied to other problems.

\citet{Liu2021} developed a custom numerical procedure based on the method of characteristics to solve the system of equations \citep{Liu2021}. This addresses issues of instability, and Liu demonstrates solutions in singularly-connected and multiply-connected domains. However, the method is limited to cases with geometric boundaries that are aligned with the characteristic curves of the equation, imposing restrictions on the traction boundary conditions that may be applied. Accuracy and computation time of this proposed approach needs to be quantitatively assessed, as well.

Finite volume (FV) methods are well suited to hyperbolic systems with irregular domains and unstructured meshes \citep{Wang2002a}. This research field is relatively mature, and the discretization approach is regularly applied to linear and non-linear problems in various applications \citep{Chai1994,Jasak1996,Mingham1998,Hermeline2000}. In these applications, one utilizes the time dependence to give a stable solution procedure. However, \cref{eq:system0} is not time dependent, and there is no readily available FV discretization that can be directly applied, hence, this motivates developing such a method. 

In order to develop a FV discretization, it is important to understand the structure of the equation in the case of a discontinuity. This includes continuous problems because the domain will be discretized into volume elements with variables and coefficients that discontinuously change. Once the behavior of the governing equation in the presence of these discontinuities is deduced, a discretization can be formulated. Interestingly, reflection phenomena arise at the discontinuities as the equations are second order and hyperbolic. 

Unlike when solving the forward problem, one cannot directly apply the known traction boundary conditions when solving \cref{eq:system0}. For example, at a location where all the characteristic lines are leaving the domain, the traction is entirely determined as part of the solution \citep{Cameron2021}. Instead of directly specifying the traction, one must translate knowledge of the traction and characteristics into the boundary conditions. In this article we analyze the general case and determine the equations that govern this.

This paper is structured as follows. In \cref{sec:theory} we present analysis addressing the characteristic curves, discontinuities, reflection, and boundary conditions. In \cref{sec:analyticalexamples} we give analytical examples of an idealized shear band and the Riemann problem, a common theoretical problem for hyperbolic partial differential equations. Finally, in \cref{sec:fv_formulation} we use the results to develop a FV discretization which is validated against simulated data.

\section{Theory}
\label{sec:theory}
\subsection{Background}
\label{ssec:background}
We review the governing equations and core assumptions which are presented in \citet{Cameron2021} or alternatively \citet{Liu2021}. Consider the simplest case of infinitesimal isotropic elastic deformation that is potentially non-linear. The spectral decomposition for the strain, $\mathbf{E}$, and the Cauchy stress, $\mathbf{T}$, is given by:
\begin{equation}
    \mathbf{E} = \mathbf{Q_E\Lambda_EQ_E^\top },\qquad \mathbf{T} = \mathbf{Q_T\Lambda_TQ_T^\top }.
\end{equation}
The key assumption is that $\bQ_\bT=\bQ_\bE$, which corresponds to the two tensors sharing the same eigenvectors. This is consistent with isotropic elasticity including non-linear hyperelastic models. We can therefore deduce the following linear relationship for $\bT$
\begin{equation}
    [\mathbf{Q^\top _ETQ_E}]_{ij} = 0 \qquad \mathrm{for}\quad i\neq j.
    \label{eq:codir_small_elastic}
\end{equation}
Intuitively, this corresponds to observing the stress tensor using a coordinate system aligned with the principal directions of the strain and then specifying that the shear components of the stress in this coordinate system are zero. We assume that $\bQ_\bE$ and other kinematic variables are known from experimental full-field deformation measurement, parameters such as $\mathbf{b}$ and $\rho$ are known, and the boundary conditions are specified. $\bT$ will be the only unknown. When \cref{eq:codir_small_elastic} is combined with force balance, this gives a total of six equations and six unknowns (\cref{eq:system0}). Hence, the solution is uniquely specified (with some exceptions discussed in \cref{ssec:characteristics}).   

The traction boundary conditions are required over a subset of the domain boundary - the areas of the boundary with incoming characteristic curves (discussed later in this subsection). These boundary conditions are known to be zero at free boundaries, and other boundary information can be obtained through careful experimental design - either measuring the traction at the boundaries or designing the problem in such a way that accurate assumptions can be made. In the future, it may also be possible to calculate the boundary conditions through optimization procedures and imposing various consistency requirements on the solution. 

The prior argument used to determine the principal direction of stress can be applied to plastic deformation if one makes two key assumptions. First, that $\mathbf{\dot{E}}^e<<\mathbf{\dot{E}}^p$. This may be practical in cases where material elements are undergoing large strains (much larger than the yield strain), although issues arise in heterogeneous deformations where different regions are elastically and elasto-plastically deforming. Second, that the principal directions of stress are aligned with $\dot{\bE}\approx\mathbf{\dot{E}}^p$. This is commonly used in constitutive equations modeling isotropic plastic deformation, such as those based on the co-directionality hypotheses. If these assumptions are met, one can determine the principal directions directly from $\dot{\bE}$.

These equations can be easily generalized to the case of finite deformation. Before proceeding, we state the notation used. The position of a point in the reference body is given by $\mathbf{X}$. The smooth one to one mapping from the reference body to the deformed body is given by $\mathbf{x}=\bar{\bx}(\mathbf{X},t)$. The deformation gradient, velocity, and velocity gradient are then given by
\begin{equation}
    \mathbf{F} = \mathrm{Grad}(\bx), \qquad \mathbf{v} = \dot{\bx}, \qquad \mathbf{L} = \mathrm{grad}(\mathbf{v}) = \dot{\mathbf{F}}\mathbf{F^{-1}}.
\end{equation}
The right and left Cauchy-Green deformation tensors are
\begin{equation}
    \mathbf{C} = \mathbf{U^2} = \mathbf{F^\top F}, \qquad \mathbf{B} = \mathbf{V^2} =  \mathbf{FF^\top }.
\end{equation}
$\mathbf{L}$ can be decomposed into its symmetric and skew parts
\begin{equation}
    \mathbf{D} = \frac{1}{2}[\mathbf{L}+\mathbf{L^\top }], \qquad
    \mathbf{W} = \frac{1}{2}[\mathbf{L}-\mathbf{L^\top }].
\end{equation}

In the case of isotropic elastic finite deformation, the principal directions of stress align with the left Cauchy-Green deformation tensor $\bB$. And in the case of isotropic finite plastic deformation, where the elastic strain is negligible, the principal directions align with $\bD$. Again, these assumptions are consistent with established constitutive equations for large deformation elasticity and plasticity (e.g. \cite{Mooney1940}). Furthermore, the alignment with $\mathbf{\dot{E}}$ or $\bD$ is consistent with viscous deformation in infinitesimal or finite deformation case respectively, so the same approach can be applied to experimental fluid mechanics. Thus we have
\begin{equation}
    \begin{aligned} 
      [\mathbf{Q^\top _ETQ_E}]_{ij} = 0 \quad \textrm{for}\quad i\neq j \qquad & \textrm{infinitesimal elastic deformation}, \\ 
      [\mathbf{Q^\top _{\dot{E}} T Q_{\dot{E}}}  ]_{ij} = 0 \quad \textrm{for}\quad i\neq j \qquad & \textrm{infinitesimal plastic and/or viscous deformation}, \\ [\mathbf{Q^\top _B TQ_B}]_{ij} = 0 \quad \textrm{for}\quad i\neq j \qquad & \textrm{finite elastic deformation}, \\
      [\mathbf{Q^\top _DTQ_D}]_{ij} = 0 \quad \textrm{for}\quad i\neq j \qquad & \textrm{finite plastic and/or viscous deformation}.
   \end{aligned}
   \label{eq:temp014}
\end{equation}
These are all special cases of \cref{eq:system0b}. This system equations is linear, hyperbolic, and variable coefficient as $\mathbf{Q}$ will vary as a function of position. The boundary conditions will be discussed in \cref{ssec:boundary_conditions}. Note that the equation and variables presented correspond to the deformed reference frame. Note also that for the remainder of this paper we will assume $\bQ$, or equivalently the eigenvectors $\{\bq^{(i)}|i=1,2,3\}$, are known and we will not specify how they are obtained, except for the case studies presented in \cref{ssec:fvvalidation}. This system of equations does not apply to cases with repeated stress eigenvalues because $\bQ$ will not be unique. For the remainder of this paper, we will assume that the eigenvalues are unique, though this limitation will likely be addressed in future work. 

The two-dimensional form of this equation is important because full-field deformation data is often only obtainable in two dimensions. Here, it can be shown that \cref{eq:temp014} becomes equivalent to:
\begin{equation}
    \frac{T_{xx}-T_{yy}}{T_{xy}} = \frac{E_{xx}-E_{yy}}{E_{xy}}
\end{equation}
for the infintesimal elastic case where \bT shares eigenvectors with \bE (equivalent expressions exist in the the plastic deformation and/or finite deformation cases). This can be rearanged to a linear equation in \bT:
\begin{equation}
    T_{xx}-T_{yy}-cT_{xy} = 0,
    \label{eq:codir2}
\end{equation}
with the known variable:
\begin{equation}
    c = \frac{E_{xx}-E_{yy}}{E_{xy}}.
    \label{eq:c}
\end{equation}
\cref{eq:codir2} is valid in the case of plane stress, or any other stress state where the out-of-plane shear stress components are zero (such as an isotropic material subject to plane strain).

\begin{figure}[ht!]
\centering
\includegraphics[width=0.66\textwidth]{{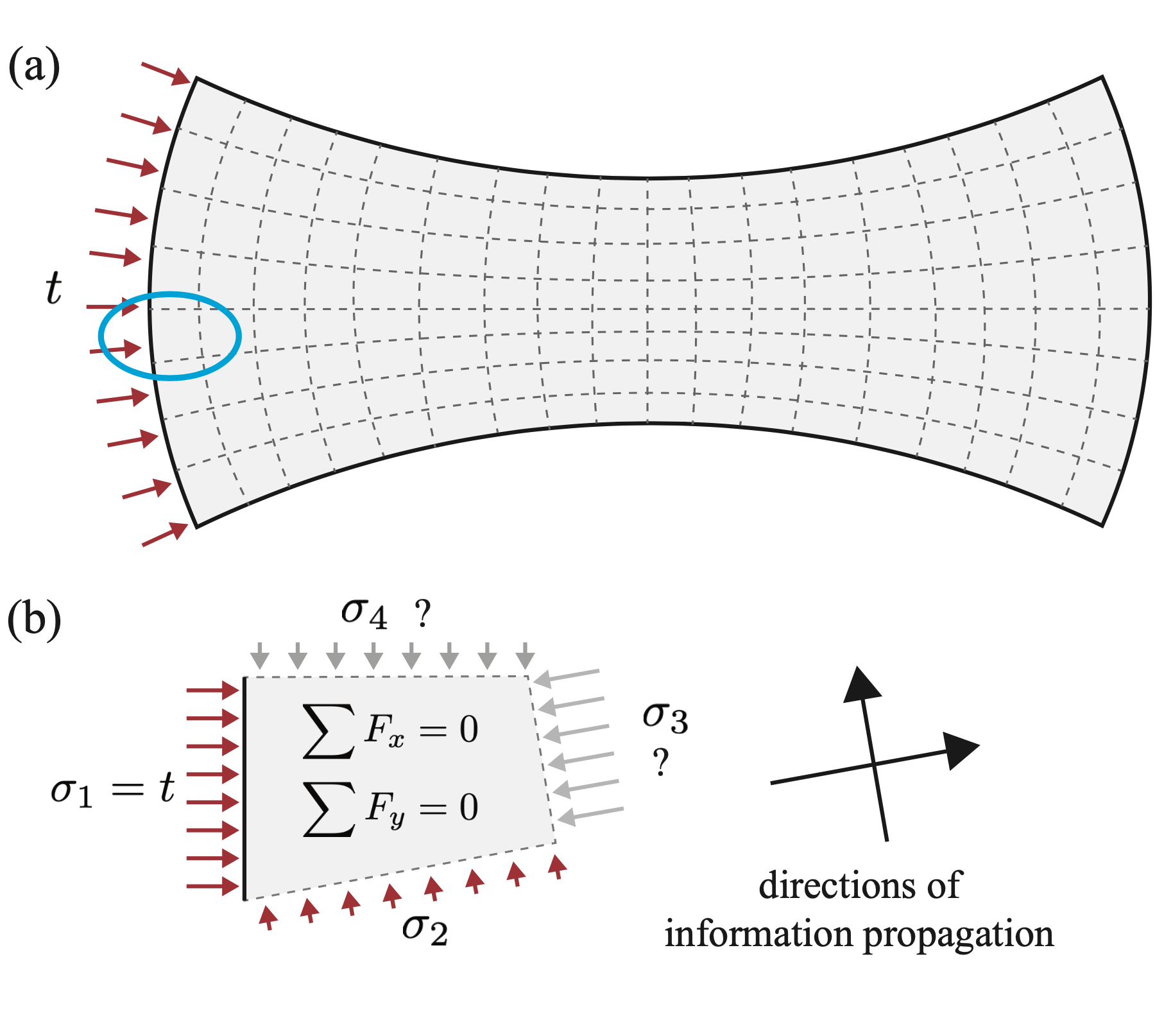}}
\caption{(a) A quasi-static deformation in two dimensions with the boundary conditions specified on one side, and a traction free boundary on the top and bottom. The dotted lines show the characteristic curves, which are aligned with the principal directions of stress at each point (see text). (b) A material element with boundaries defined by the characteristic curves. One can use force balance to compute two of the stresses from two other stresses. When this computation is conducted, information propagates from left to right and bottom to top along the characteristic curves. }
\label{fig:concept}
\end{figure}

We give a brief introduction to the characteristic curves to give the reader intuition that is important for the remainder of the paper. Consider \cref{fig:concept} which depicts a deformation in two dimensions where the principal directions of the stress are known. Also, we assume the traction boundary conditions are known. From knowledge of the principal directions, one can form a grid of curved lines aligned with the principal directions at each point. These curves, referred to as characteristic curves, will always intersect at right angles because the principal directions of \bT are orthogonal. Consider the material element shown in \cref{fig:concept}b. The boundaries of the element are comprised of these characteristic curves, hence, the boundaries are aligned with the principal directions of \bT and no shear force can be exerted across the boundary, only normal force. If we know the magnitude of the stresses on the left and bottom sides ($\sigma_1$ and $\sigma_2$), we can use force balance in the $x$ and $y$ direction to compute the stresses on the right and top sides ($\sigma_3$ and $\sigma_4$). To compute the stress field throughout the entire domain, we can start on the lower left corner of the domain, and proceed by moving upward and rightward. In the limit of small material elements, the direction of computation will be along the characteristic curves and one can think of this as information propagating along these characteristics. One interesting feature is that the boundary conditions on the top and right side will be computed as part of the solution, rather than specified in the problem (as is the case with standard FE simulations).

\subsection{Characteristic curves and principal directions}
\label{ssec:characteristics}
Before discussing discontinuities and reflections, it is important to mathematically show that the characteristic curves are indeed aligned with the principal directions of the stress. Previously, this has simply been argued on an intuitive basis \citep{Liu2021,Cameron2021}.

To show this, we carry out the following procedure. First, we combine the system of differential equations and algebraic equations into one system of differential equations. Second, we diagonalize this system and determine the eigenvalues, which in turn specify the characteristic directions. Third, we compare these characteristic directions to the principal directions of the stress to show they are the same. We present equations and arguments in two dimensions as the formulas are more concise and the relevant physics is the same as the three-dimensional case.

We start with the following equations
\begin{gequation}
    \label{eq:force1}
    \dx{\Txx}+\dy{\Txy} = 0,\\
    \dx{\Txy}+\dy{\Tyy} = 0,\\
\end{gequation}
and
\begin{equation}
    \Txx - \Tyy - c\Txy = 0.
    \label{eq:allignment1}
\end{equation}

One can substitute for $T_{xx}$ in \cref{eq:force1} using \cref{eq:allignment1}
\begin{equation}
    \begin{gathered}
    \dx{\Tyy}+c\dx{\Txy}+\dy{\Txy}+\Txy\dx{c}=0,\\
    \dx{\Txy}+\dy{\Tyy}=0.
    \end{gathered}
    \label{eq:governing1}
\end{equation}
This can be written in matrix form:
\begin{equation}
    \dx{\bu} + \bA\dy{\bu}+\mathbf{B}\bu = 0,
\end{equation}
where,
\begin{equation}
    \bu = 
    \begin{bmatrix}
    \Tyy \\
    \Txy
    \end{bmatrix}, \qquad
    \bA = 
    \begin{bmatrix}
    0 & 1\\
    1 & c
    \end{bmatrix}, \qquad
    \mathbf{B} = 
    \begin{bmatrix}
    0 & 0\\
    0 & \dx{c}
    \end{bmatrix}.
\end{equation}

According to standard theory on hyperbolic partial differential equations (e.g. see \cite{LeVeque2002}), the eigenvalues of \bA, $\{\lambda_\bA^{(\alpha)}|\alpha = 1,2\}$, determine the characteristic curves of the equation, and the eigenvectors of \bA, $\{\bq_\bA^{(\alpha)}|\alpha = 1,2\}$, give the information that propagates along these characteristics. Note that we use Greek letters to indicate the index when it takes values in the set $\{1,2\}$. One may express the unknown $\bu$ as a linear combination of the eigenvectors of $\mathbf{A}$
\begin{equation}
    \mathbf{u} = a^{(1)}\bq_\bA^{(1)} + a^{(2)}\bq_\bA^{(2)}.
    \label{eq:temp015}
\end{equation}
Here, the $a^{(1)}\bq_\bA^{(1)}$ component of the stress will propagate along the characteristic determined by $\lambda_\bA^{(1)}$. Similarly, the $a^{(2)}\bq_\bA^{(2)}$ component will propagate along the characteristic determined by $\lambda_\bA^{(2)}$. We have
\begin{equation}
    \begin{gathered}
    \lambda^{(1)}_\bA = (c-\sqrt{c^2+4})/2, \qquad \lambda^{(2)}_\bA  = (c+\sqrt{c^2+4})/2,\\
    \bq_\bA^{(1)}  = \begin{bmatrix}
    -(c+\sqrt{c^2+4})/2\\
    1
    \end{bmatrix}
    ,\qquad
    \bq_\bA^{(2)}  = \begin{bmatrix}
    (-c+\sqrt{c^2+4})/2\\
    1
    \end{bmatrix}
    \end{gathered}.
\end{equation}
The vectors $\mathbf{v}^{(\alpha)}$ corresponding to the tangents to the characteristic curves can then be computed using the equation defining the $\alpha$ characteristic curve, $dx= \lambda^{(\alpha)}_\mathbf{A} dy $. This gives
\begin{equation}
    \mathbf{v}^{(\alpha)}= \begin{bmatrix}
    \lambda_\mathbf{A}^{(\alpha)} \\
    1
    \end{bmatrix}.
\end{equation}
Hence we have
\begin{equation}
    \mathbf{v}^{(1)}= \begin{bmatrix}
    (c-\sqrt{c^2+4})/2 \\
    1
    \end{bmatrix}, \qquad
    \mathbf{v}^{(2)}= \begin{bmatrix}
    (c+\sqrt{c^2+4})/2 \\
    1
    \end{bmatrix},
    \label{eq:temp019}
\end{equation}
where these vectors are defined up to a multiplicative constant. This gives the spatial direction along which information propagates throughout the material. 

We now consider the components of stress that propagate along each characteristic curve, which can be computed from $a^{(1)}\bq_\bA^{(1)}$ and $a^{(2)}\bq_\bA^{(2)}$ respectively. We label these $\bT^{(1)}$ and $\bT^{(2)}$. Due to \cref{eq:temp015} and the definition of $\bu$ we have
\begin{equation}
    \bT = \bT^{(1)}+\bT^{(2)}
    \label{eq:temp016},
\end{equation}
where $\bT$ is the stress tensor in matrix form (in contrast to the vector form $\mathbf{u}$). First we  consider the stress matrix corresponding to the component of $\bu$ that propagates along $\alpha =1$ characteristic curve, given by $a^{(1)}\bq_\bA^{(1)}$. Using the expression for $\bq_\bA^{(1)}$ we have $\Tyy^{(1)} = -a^{(1)}(c+\sqrt{c^2+4})/2$, and $\Txy^{(1)} = a^{(1)}$. \cref{eq:allignment1} gives the $\Txx^{(1)}$ component. Hence, we have
\begin{equation}
    \begin{gathered}
    \bT^{(1)} = a^{(1)}\begin{bmatrix}
    (c-\sqrt{c^2+4})/2 & 1\\
    1 &  -(c+\sqrt{c^2+4})/2
    \end{bmatrix}.
    \end{gathered}
\end{equation}
This is a tensor with matrix rank one, as one of the eigenvalues are zero (this should not be confused with a vector, or first-order tensor, which is often referred to as a rank one tensor). The non-zero eigenvalue and corresponding eigenvector of $\bT^{(1)}$ is
\begin{equation}
\label{eq:temp01}
    \begin{gathered}
    \lambda_{\bT (1)} = -a^{(1)}\sqrt{c^2+4},\\
    \bq_{\bT (1)} = \begin{bmatrix}
    (c-\sqrt{c^2+4})/2\\
    1
    \end{bmatrix},
    \end{gathered}
\end{equation}
where the $(1)$ is in the subscript position to indicate that it corresponds to the $\bT^{(1)}$ matrix. One can compare this expression with \cref{eq:temp019} to conclude that the eigenvector of the stress matrix $\bT^{(1)}$ is the same as $\bv^{(1)}$, and gives the characteristic direction for the $\bT^{(1)}$ component of the stress. 

One can repeat the same exercise for the $\alpha=2$ characteristic curve
\begin{gequation}
\bT^{(2)} = a^{(2)}\begin{bmat}
 (c+\sqrt{c^2+4})/2 & 1\\
 1 & -(c-\sqrt{c^2+4})/2
\end{bmat},
\end{gequation}
\begin{gequation}
\label{eq:temp02}
    \lambda_{\bT (2)} = -a^{(2)}\sqrt{c^2+4},\\
    \bq_{\bT (2)} = \begin{bmat}
    -(c-\sqrt{c^2+4})/2\\
    1
    \end{bmat}.
\end{gequation}
Again, this shows that the principal direction of the stress is coaxial with the respective characteristic curve. 

One can substitute the spectral decompositions for each $\bT^{(\alpha)}$ into \cref{eq:temp016}, noting that each $\bT^{(\alpha)}$ is matrix rank one
\begin{equation}
    \bT = \lambda_{\bT (1)}\bq_{\bT (1)}\bq^\top _{\bT (1)} + \lambda_{\bT (2)}\bq_{\bT (2)}\bq^\top _{\bT (2)}.
    \label{eq:temp018}
\end{equation}
The spectral decomposition for $\bT$ in two dimensions can be expressed as 
\begin{equation}
    \bT = \sigma^{(1)}\bq^{(1)}\bq^{(1)\top}+\sigma^{(2)}\bq^{(2)}\bq^{(2)\top},
    \label{eq:temp017}
\end{equation}
where $\{\sigma^{(\alpha)}\}$ are the principal stresses and $\{\bq^{(\alpha)}\}$ (without the subscript) are the principal directions. 

Comparing \cref{eq:temp018,eq:temp017}, one can take $ \lambda_{\bT (\alpha)} = \sigma^{(\alpha)}$ and $\bq_{\bT (\alpha)}=\bq^{(\alpha)}$. Hence, the stress can simply be decomposed into two matrix-rank one tensors using the spectral decomposition, and each tensor will propagate along the direction of its non-zero principal direction. We can interpret this as the principal stresses simply propagating along their respective principal directions. This is the key result of the argument just made and allows one to make a clear physical interpretation of the characteristic curves.

We note that the same argument can be made in three dimensions, leading to the same conclusion: that the principal directions of stress are aligned with the respective characteristics of the governing equation. Hence, it will frequently be convenient to decompose $\bT$ into its eigenvalues and eigenvectors
\begin{equation}
    \bT = \sum_i\sigma^{(i)}\bq^{(i)}\bq^{(i)\top}.
    \label{eq:Tdecomposition}
\end{equation}

\begin{figure}[ht!]
\centering
\includegraphics[width=1\textwidth]{{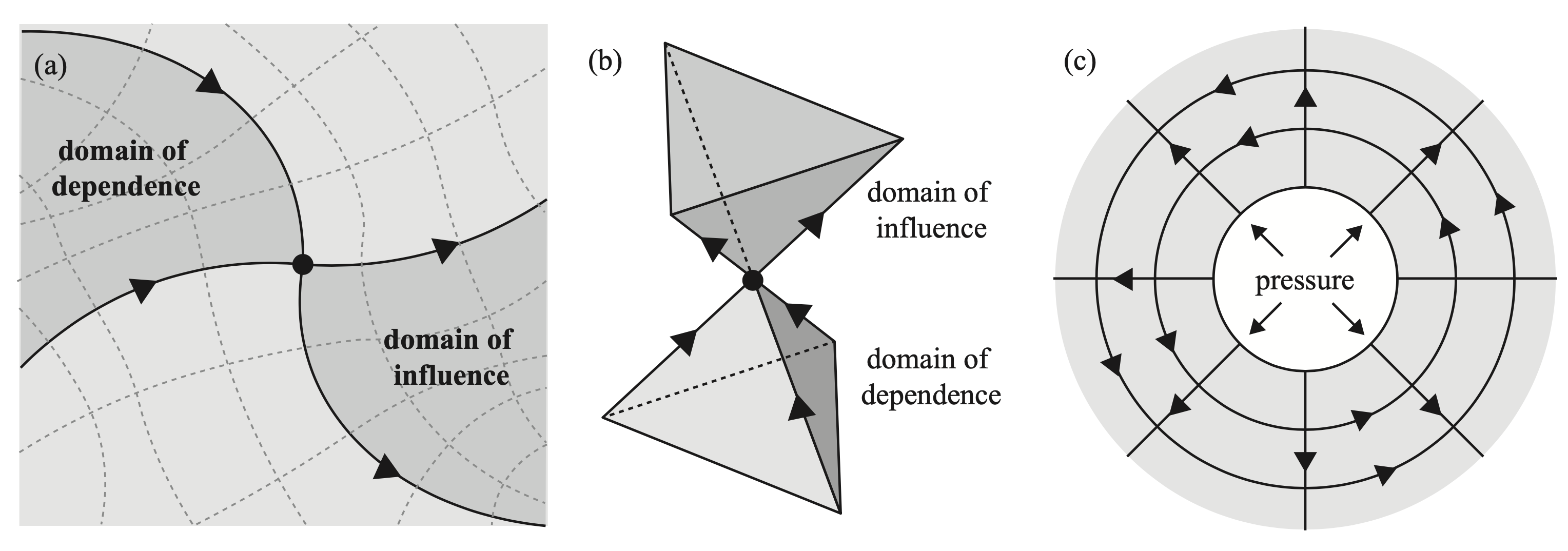}}
\caption{(a) The domain of dependence and influence in two dimensions where the characteristic directions are varying with position. (b) The domain of dependence and influence in three dimensions where the characteristic directions are constant. (c) An example where the solution is non-unique as the characteristic lines form closed loops. This example shows a rotationally symmetric problem where a hole exerts a constant pressure on the domain boundary. }
\label{fig:equation_structure}
\end{figure}

We make several remarks:
\begin{itemize}
    \item One must assign the direction of information propagation along each characteristic curve when solving the system of equations. This gives $2^2=4$ possibilities in two dimensions and $2^3=8$ possibilities in three dimensions, one binary choice for each characteristic curve. The choice of the direction of information propagation is intrinsically linked to which boundary conditions are specified, as the principal stresses can only be specified along the characteristics entering the domain.
    \item The stress at a particular point is only determined by the stress in the domain of dependence (\cref{fig:equation_structure}).  The stress at a particular point influences stresses only in its domain of influence. We refer to \textit{upstream} and \textit{downstream} as positions along characteristic curves reverse to the direction of information propagation and along the direction of information propagation respectively. Points in the domain of dependence can be reached by following any combination of characteristic lines upstream from the point of interest. Similarly, the points in the domain of influence can be found by following any combination of characteristics in the downstream direction. 
    \item The characteristic curves are always perpendicular to one another when they intersect. In two dimensions, the variable coefficient $c$ controls the orientation of the characteristic curves, however, in three dimensions, three independent variables are required. We will often use  $\{\bq^{(i)}\}$ in two or three dimensions to specify the orientation of the characteristic lines. In some parts of the paper, it will be more concise to refer to the principal direction orientation angle with respect to the $x$ axis, $\theta$. The other principal direction will be oriented at $\theta+\pi/2$. We have the expression
    \begin{equation}
        \theta = \tan^{-1}\left(\frac{1}{c+\sqrt{1+c^2}}\right)
    \end{equation}
    \item There are several forms of the governing equation which may be useful in different applications. For example, one equation form can be obtained in two dimensions using the stress Airy function
    \begin{equation}
        \frac{\partial \psi}{\partial y^2} -  \frac{\partial \psi}{\partial x^2} + c\frac{\partial \psi}{\partial x \partial y} = 0.
    \end{equation}
    This can be decomposed into a system of first-order equations
    \begin{equation}
    \begin{gathered}
    \frac{\partial \psi}{\partial y} + \frac{1}{2}(c+ \sqrt{c^2+4})\frac{\partial \psi}{\partial x}=0,\\
    \frac{\partial \psi}{\partial y} + \frac{1}{2}(c-\sqrt{c^2+4})\frac{\partial \psi}{\partial x}=0,
    \end{gathered}
\end{equation}
where we assume there is no body force or acceleration term. This particular form may be useful in some applications as there is no source term.
\item While in many cases the solution is deterministic, non-uniqueness arises when the characteristic curves form closed loops. For example, the case of a rotationally symmetric expansion depicted in \cref{fig:equation_structure}c. This will likely the subject of future work.
\end{itemize}

\subsection{Discontinuities}
\label{ssec:strain_discontinuity}
Next, we consider a simple discontinuity in the principal directions of the stress. This could arise from discontinuities in the observed strain or strain rate field, however, it also arises in the finite volume discretization process even when the strain field is continuous. Hence, it is necessary to understand prior to developing the discretization procedure. We will show how discontinuities give rise to reflection and transmission phenomena. For simplicity, we assume the material is in static equilibrium and that the problem is two dimensional (we comment at the end of this subsection on the three-dimensional case). We also note that one can solve these problems trivially by balancing traction across the interface and utilizing \cref{eq:codir2}, however, it is insightful to show how the same solution arises directly from the system of partial differential equations. We will then utilize the simpler traction balance approach for the remainder of the paper. 

\begin{figure}[ht!]
\centering
\includegraphics[width=1\textwidth]{{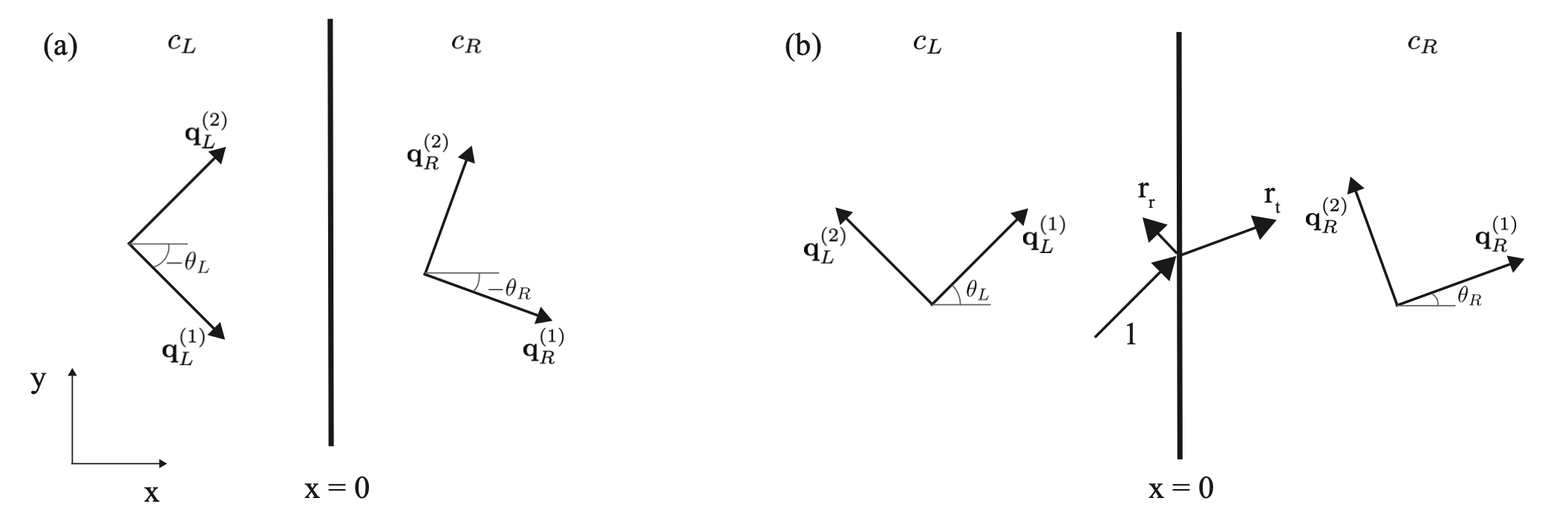}}
\caption{Solving the equations over a discontinuity for different characteristic directions. On the left and right hand side of the discontinuity the principal directions, or characteristics, have different orientations. (a) The stress on one side of the discontinuity is known. (b) One principal stress is known on each side. This latter case leads to reflection and transmission of the principal stresses. }
\label{fig:strain_discontinuity}
\end{figure}

First, consider the problem depicted in \cref{fig:strain_discontinuity}a. Here, we assume the stress is known for $x<0$, and must be computed for $x>0$. The coefficient $c$ is given by
\begin{equation}
    c = \left.
  \begin{cases}
    c_L, & \text{if } x \leq 0, \\
    c_R, & \text{if } x > 0, \\
  \end{cases}\right.
\end{equation}
where the subscripts $L$ and $R$ correspond to left and right respectively. The unit vectors $\{\bq^{(\alpha)}_L\}$ and $\{\bq^{(\alpha)}_R\}$ are determined by $c_L$ and $c_R$ respectively using \cref{eq:temp01,eq:temp02}.  Here the solution will not depend on $y$, hence, we eliminate the $y$ derivatives from \cref{eq:governing1} giving
\begin{gequation}
    \dx{\Tyy}+c\dx{\Txy}+\Txy\dx{c}=0,\\
    \dx{\Txy}=0.
\end{gequation}

This directly gives:
\begin{equation}
\label{eq:temp04}
    T_{xyL}=T_{xyR}=\Txy,
\end{equation}
and 
\begin{equation}
\label{eq:temp03}
    \dx{\Tyy}+\Txy\dx{c}=0.
\end{equation}
We have the expression for $\dx{c}$
\begin{equation}
    \dx{c} = (c_R-c_L)\delta(x),
\end{equation}
where $\delta(x)$ is the Dirac delta function. Substituting this into \cref{eq:temp03} then integrating we have:
\begin{gequation}
\label{eq:temp05}
    \dx{\Tyy} = -\Txy(c_R-c_L)\delta(x),\\
    T_{yyR}-T_{yyL} = -\Txy(c_R-c_L),\\
    T_{yyR} = \Txy(c_L-c_R)+T_{yyL}.
\end{gequation}
Finally, by calculating $T_{xxL}$ and $T_{xxR}$ using \cref{eq:allignment1} one can obtain:
\begin{equation}
\label{eq:temp06}
    T_{xxL} = T_{xxR}.
\end{equation}

Together, \cref{eq:temp04,eq:temp05,eq:temp06} give expressions for the stress where $x>0$ in terms of the stress where $x\leq0$. This solution is simply equivalent to balancing the traction across the discontinuity and specifying the principal directions.

One can repeat the same exercise for a $y=\mathrm{constant}$ discontinuity, where force balance plus the constraint on the principal directions is sufficient to determine the unknown stress. We can express the general formula relating the stress on each side of a discontinuity as
\begin{equation}
    \bT_A \bn = \bT_B\bn,
\end{equation}
where $\bn$ is the vector normal to the discontinuity, and A and B correspond to two regions on each side of the discontinuity. In two or three dimensions, this can be expressed in the following equations relating $\{\sigma^{(i)}_A,\sigma^{(i)}_B\}$
\begin{equation}
    \sum_i\sigma_A^{(i)}\bq_A^{(i)}\bq^{(i)\top}_A\bn = \sum_i\sigma_B^{(i)}\bq_B^{(i)}\bq^{(i)\top}_B\bn.
    \label{eq:temp020}
\end{equation}

It may not be true that the stress is fully determined on one side of the discontinuity and unknown on the other: it may be partially determined on each side. For the general case in two dimensions, two characteristics are incoming to the boundary, and two are outgoing from the boundary. In three dimensions, there are three characteristics incoming and three outgoing. Given that $\{\sigma^{(i)}\}$ along the incoming characteristics is assumed known, $\{\sigma^{(i)}\}$ along the outgoing characteristics can be computed using \cref{eq:temp020}.

We briefly consider the case where characteristics are parallel to the discontinuity. In two dimensions, where one characteristic curve on one side of the discontinuity is parallel to the discontinuity, two independent force balance equations will remain. Therefore, there should be two characteristics outgoing from the discontinuity and one incoming so the solution is unique. In three dimensions, there are several possibilities which depend on the number of characteristics parallel to the discontinuity. \cref{eq:temp020} reduces to a different number of equations depending on the scenario. For example, in the case of two characteristics on each side being parallel to the boundary, \cref{eq:temp020} reduces to one equation, so there should be one outgoing characteristic. 

One can understand this analysis in terms of the "reflection" and "transmission" of the principal stresses. In general, reflection and transmission phenomena arise whenever the system is governed by a set of partial differential equations that are linear, hyperbolic, second order or higher, variable coefficient, and discontinuous in the variable coefficient(s). To give the reader intuition for this, we consider the analogy of a sound wave. The mathematics governing this refection is very similar to the mathematics governing the behavior of the principal stresses at the discontinuity. Specifically, a discontinuity in the principal directions of stress is analogous to the discontinuity in sound wave speed, as the wave speed is essentially a "direction" in the space-time domain. In addition, the principal stresses themselves are analogous to the amplitudes of sound waves propagating in different directions. When a sound wave encounters a wave speed discontinuity, a component reflects off the discontinuity and another transmits through it, and the amplitudes of each component are determined by the respective wave speeds and Newton's second law.  In the principal stress case, the principal stress propagating along the direction incoming toward the discontinuity is converted into two principal stresses propagating along directions outgoing from the discontinuity. We can say that a component of the principal stress is "reflected" and a component is "transmitted". (The reader can refer to the example in \cref{ssec:shearband} which may give further intuition).

Consider a principal stress $\sigma^{(1)}_L$ on an incoming characteristic, $\bq^{(1)}_L$, as shown in \cref{fig:strain_discontinuity}b, where we assume $\sigma^{(2)}_R=0$. The incoming principal stress, $\sigma^{(1)}_L$, will have a reflected component along $\bq^{(2)}_L$ of magnitude $\sigma^{(2)}_L = r_r\sigma^{(1)}_L$, and a transmitted component along $\bq^{(2)}_R$ of magnitude $\sigma^{(2)}_R= r_t\sigma^{(1)}_L$. To calculate $r_r$ and $r_t$ we take $\sigma^{(1)}_L=1$ and use \cref{eq:temp020}. One arrives two equations and two unknowns that can be solved
\begin{equation}
    r_t\bq_R^{(1)}\bq^{(1)\top}_R\bn-r_r\bq^{(2)}_L\bq^{(2)\top}_L\bn=\bq^{(1)}_L\bq^{(1)\top}_L\bn.
\end{equation}
Hence, $r_r$ and $r_t$ are determined by the principal directions $\{\bq^{(\alpha)}_L,\bq^{(\alpha)}_R\}$, and boundary normal $\bn$. Substituting for $\{\bq^{(\alpha)}_L,\bq^{(\alpha)}_R\}$, and taking $\bn$ to be in the $x$ direction gives
\begin{equation}
    r_t = \frac{\cos(\theta_L)}{\cos(\theta_L-\theta_R)\cos(\theta_R)}, \qquad
    r_r = \frac{\tan(\theta_L-\theta_R)}{\tan(\theta_L)},
\end{equation}
where we express the answer using $\theta_L$ and $\theta_R$ defined in \cref{ssec:characteristics}). Finally, when considering the case of two incoming principal stresses, one can simply apply the principal of superposition: the $\sigma^{(i)}$ on an outgoing characteristic can be expressed as the sum of values each arising from the reflection or transmission of an incoming characteristic. This same approach can be applied directly to the three-dimensional case.

\subsection{Boundary conditions}
\label{ssec:boundary_conditions}
This section discusses how knowledge of the traction at the boundary can be translated to specify the boundary condition. Specifically, we must specify the principal stresses along the characteristics entering the domain, and not specify the principal stresses along characteristics leaving the domain. 

Here, we utilize the results from the prior subsections showing the characteristics are aligned with the principal directions of the stress, and solving the equations at a discontinuity simply corresponds to balancing the traction. In general, $\{\sigma^{(i)}\}$ must be specified on the characteristics entering the domain, $\{\bq^{(ei)}\}$, but cannot be specified on those leaving the domain, $\{\bq^{(li)}\}$. If $\bn$ is the unit outward normal of the domain boundary, we have $\bq^{(li)\top}\bn>0$ and $\bq^{(ei)\top}\bn<0$. We have the additional special case of characteristics aligned with the boundary, $\{\bq^{(ai)}\}$, where $\bq^{(ai)\top}\bn=0$. We have $\{\bq^{(i)}\} = \{\bq^{(ei)}\} \cup \{\bq^{(li)}\}\cup \{\bq^{(ai)}\}$. There will be $n$ characteristics at each point, where $n=3$ in three dimensions and $n=2$ in two dimensions. Let $n^e$ denote the number of characteristics entering the domain, $n^l$ denote the number of characteristics leaving the domain, and $n^a$ denote the number of characteristics aligned with the domain boundary such that $n^e+n^l+n^a=n$. 

In general, we will utilize the equation
\begin{equation}
    \sum_i\sigma^{(i)}\bq^{(i)}\bq^{(i)\top}\bn = \bt,
    \label{eq:temp021}
\end{equation}
where \bt is the traction at the boundary.  In the straightforward case where all characteristics are entering the domain ($n^e=n$), \cref{eq:temp021} gives three equations and three unknowns which can be solved for $\{\sigma^{(i)}\}$. Similarly, in the case where all characteristics are leaving the domain ($n^l=n$), $\bt$ is entirely determined from the values in the domain so no boundary traction can be specified.

In the case where $n^a\neq0$, the term(s) containing $\bq^{(ai)\top}\bn$ in \cref{eq:temp021} become zero eliminating the unknown(s) $\sigma^{(ai)}$ from the equation. Hence, the system of equations will become over-determined for a general $\bt$. In order to resolve this issue, one must specify a $\bt$ such that $\bq^{(ai)\top}\bt=0$. Physically, this corresponds to the traction on a surface having no shear component when a principal direction of the stress is orthogonal to the surface.

For any value of $n^e$, one can specify $n^e$ components of the traction. In the case where $0<n^e<n$, one can also choose which components of the traction are specified. Let $\{\mathbf{d}^{(j)}\}$ be any orthonormal basis spanning the two- or three-dimensional space. One may choose a subset of $\{\mathbf{d}^{(j)}\}$  of size $n^e$ in which to specify the traction components. We denote this $\{\mathbf{d}^{(j)*}\}$. Taking the dot product of \cref{eq:temp021} with $\mathbf{d}^{(j)*}$ gives:
\begin{equation}
    \sum_i\sigma^{(i)}(\bd^{(j)*\top}\bq^{(i)})(\bq^{(i)\top}\bn) = \bd^{(j)*\top}\bt =t_j.
\end{equation}
This is a scalar equation involving $\{\sigma^{(i)}\}$. There will be $n^e$ equations and $n^e$ unknowns, so the equations can be solved. One is free to select any $n^e$ orthonormal vectors $\{\mathbf{d}^{(i)*}\}$, with the only restriction being that $\bd^{(i)} \neq \pm \bq^{(lj)}$, otherwise the equations for $\sigma^{(ei)}$ will be singular. Freedom to select $\{\mathbf{d}^{(i)*}\}$ allows one to:
\begin{itemize}
    \item specify boundary conditions that are non-reflective by choosing $\{\mathbf{d}^{(j)*}\} = \{\bq^{(ei)}\}$
    \item specify the orientation of $\bt$ by choosing particular $t_j=0$
    \item specify the magnitude of $\bt$ in certain directions by choosing particular $t_j$ and $\mathbf{d}^{(j)*}$.
\end{itemize}

\begin{figure}[ht!]
\centering
\includegraphics[width=0.5\textwidth]{{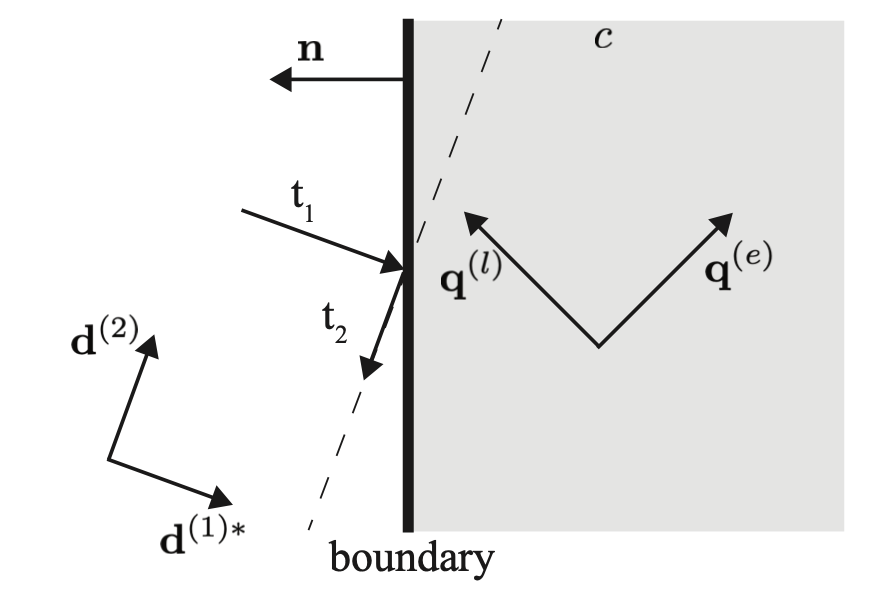}}
\caption{Determining the boundary conditions when one characteristic is entering the domain, $\bq^{(e)}$, and another is leaving, $\bq^{(l)}$. One can specify the traction components in chosen $\mathbf{d^{(1)*}}$ directions.}
\label{fig:boundary_traction}
\end{figure}

We consider an example in two dimensions of one characteristic entering the domain and one characteristic leaving the domain. 
Here, one can specify one component of the traction vector, whereas the other component is determined by the principal stress along the leaving characteristic (\cref{fig:boundary_traction}). Let $\bd^{(1)*}$ be a unit vector corresponding to the direction where one chooses to specify the traction, and $\bd^{(2)}$ correspond to the direction where $\bt$ will be determined. Let $t_1 = \bd^{(1)*\top}\bt$ and $t_2 = \bd^{(2)\top}\bt$. Let $\bq^{(e)}$ correspond to the characteristic entering the domain and $\bq^{(l)}$ correspond to the characteristic leaving the domain. One must determine $\sigma^{(e)}$, where $\sigma^{(l)}$ is known. We have
\begin{equation}
    \sigma^{(1)}\bq^{(1)}\bq^{(1)\top}\bn+\sigma^{(e)}\bq^{(e)}\bq^{(e)\top}\bn = \bt.
\end{equation}
We take the dot product of this equation with $\bd^{(1)*}$ and $\bd^{(2)}$ respectively giving
\begin{gequation}
\sigma^{(l)}\bd^{(1)*\top}\bq^{(l)}\bq^{(l)\top}\bn+\sigma^{(e)}\bd^{(1)*\top}\bq^{(e)}\bq^{(e)\top}\bn = t_1,
\label{eq:temp09}
\end{gequation}
\begin{equation}
\sigma^{(l)}\bd^{(2)\top}\bq^{(l)}\bq^{(l)\top}\bn+\sigma^{(e)}\bd^{(2)}\bq^{(e)}\bq^{(e)\top}\bn = t_2.
\label{eq:temp010}
\end{equation}
rearranging \cref{eq:temp09} gives:
\begin{equation}
    \sigma^{(e)} = \frac{t_1-\sigma^{(l)}\bd^{(l)*\top}\bq^{(l)}\bq^{(l)\top}\bn}{\bd^{(l)*\top}\bq^{(e)}\bq^{(e)\top}\bn}.
\end{equation}
This gives the principal component for the entering characteristic. One can interpret this as being comprised of two terms, one due to the specified component of traction along the boundary, and another due to reflection off the boundary from the leaving characteristic. We also obtain the expression for the unknown component of the traction by substituting this into \cref{eq:temp010}
\begin{equation}
    t_2 = \sigma^{(l)}\bd^{(2)\top}\bq^{(l)}\bq^{(l)\top}\bn + \frac{(t_1-\sigma^{(l)}\bd^{(1)*\top}\bq^{(l)}\bq^{(l)\top}\bn)\bd^{(2)\top}\bq^{(e)}}{\bd^{(1)*\top}\bq^{(e)}}.
\end{equation}
One has freedom to select $\bd^{(1)*}$ except for the restriction mentioned above. To eliminate reflection at the boundary, one may specify $\bd^{(1)*} = \bq^{(e)}$ and $\bd^{(2)}= \bq^{(l)}$. This gives
\begin{gequation}
    \sigma^{(e)} = \frac{t_1}{\bq^{(e)\top}\bn},\\
    t_2 = \sigma^{(l)}\bq^{(l)\top}\bn.
    \label{eq:bc_norefection}
\end{gequation}

We remark on a few boundaries of physical relevance. First, consider a free boundary that will have zero traction. Here, one characteristic is orthogonal to the boundary, and the other characteristic(s) will align with it. In the case where the orthogonal characteristic is entering the domain, one can specify that the normal component of the traction is zero. Second, consider a domain boundary that is in contact with a frictionless rigid surface. Here, there will also be one characteristic orthogonal to the boundary and the other characteristic(s) will be aligned with it. In the case where one or two characteristics are entering the domain, one may specify the $\bd^{(i)*}$ to be orthogonal to $\bn$ and the corresponding traction in these direction(s) to be zero, hence, the traction vector will be normal to the boundary with a magnitude that is determined instead of specified. Third, consider the case where $\bt$ is known (or assumed). One can directly apply the equations discussed, however, in the case of one characteristic entering and one leaving, one must choose the direction(s) $\{\bd^{(i)*}\}$ along which to specify the component of $\bt$.

\section{Analytical examples}
\label{sec:analyticalexamples}
\subsection{Riemann problem}
We now consider the solution to the Riemann problem in two dimensions, which is commonly used to perform computations for various systems of hyperbolic partial differential equations.

\begin{figure}[ht!]
\centering
\includegraphics[width=0.8\textwidth]{{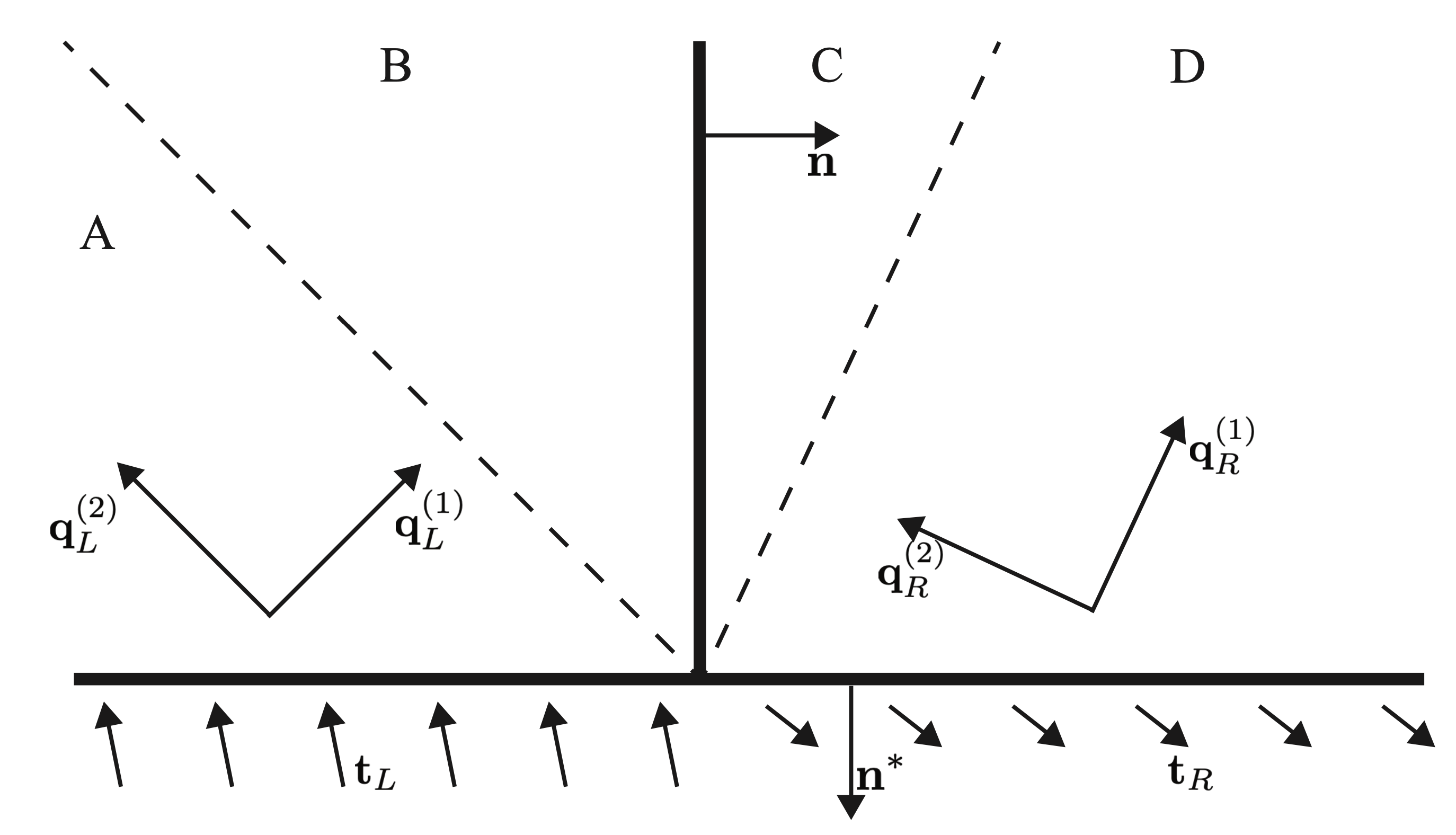}}
\caption{Schematic of the Riemann problem for the governing equations. }
\label{fig:riemann}
\end{figure}

The problem is depicted in \cref{fig:riemann}. We have a discontinuous boundary traction and a discontinuous coefficient $c$ that both change at $x=0$. We have characteristic directions $\{\bq^{(\alpha)}_L\}$ on the left side ($x\leq 0$), and $\{\bq^{(\alpha)}_R\}$ on the right side ($x>0$). Similarly the traction $\bt_L$ is applied for $x\leq 0$, and $\bt_R$ is applied for $x>0$. We specify the direction of information propagation such that the information propagates in the positive $y$ direction. We also assume the material is in static equilibrium. The stress will have four different states, one in each region A, B, C and D. Regions A-B boundary will be aligned with the $\bq^{(2)}_L$ characteristic and the C-D boundary will be aligned with the $\bq^{(1)}_R$ characteristic. The normal vector to the domain boundary is denoted $\bn^*$ and the normal vector to the B-C boundary is denoted $\bn$. We have eight unknowns: $\{\sigma^{(1)}_A,\sigma^{(2)}_A,\sigma^{(1)}_B,\sigma^{(2)}_B,\sigma^{(1)}_C,\sigma^{(2)}_C,\sigma^{(1)}_D,\sigma^{(2)}_D\}$. 

One can determine $\{\sigma^{(1)}_A,\sigma^{(2)}_A,\sigma^{(1)}_D,\sigma^{(2)}_D\}$ using force balance and the constraint on the principal directions as discussed in \cref{ssec:boundary_conditions}
\begin{gequation}
\sigma^{(1)}_A\bq_L^{(1)}\bq_L^{(1)\top}\bn^*+\sigma^{(2)}_A\bq^{(2)}_L\bq_L^{(2)\top}\bn^* = \bt_L,\\
\sigma_D^{(1)}\bq_R^{(1)}\bq_R^{(1)\top}\bn^*+\sigma_D^{(2)}\bq_R^{(2)}\bq_R^{(2)\top}\bn^* = \bt_R.
\label{eq:temp022}
\end{gequation}
One can directly solve these equations as they are two linear equations with two unknowns. Next, consider the A-B boundary. There is no change in $\{\bq^{(\alpha)}_L\}$, hence, we have
\begin{equation}
    \sigma_A^{(1)}=\sigma_B^{(1)}.
    \label{eq:temp023}
\end{equation}
Similarly, for the C-D boundary, we have
\begin{equation}
    \sigma_C^{(2)}=\sigma_D^{(2)}.
    \label{eq:temp024}
\end{equation}
Finally, force balance is used along the B-C boundary to obtain $\sigma_B^{(2)}$ and $\sigma_C^{(1)}$ (see \cref{ssec:strain_discontinuity})
\begin{equation}
\label{eq:temp011}
    \sigma^{(1)}_B\bq^{(1)}_L\bq^{(1)\top}_L\bn+\sigma^{(2)}_B\bq^{(2)}_L\bq^{(2)\top}_L\bn = \sigma^{(1)}_C\bq^{(1)}_R\bq^{(1)\top}_R\bn+\sigma^{(2)}_C\bq^{(2)}_R\bq^{(2)\top}_R\bn.
\end{equation}

The solution of this equation can be expressed using the $\theta$ notation (\cref{ssec:characteristics}) to give expressions for the unknown $\sigma_B^{(2)}$ and $\sigma_C^{(1)}$
\begin{equation}
    \sigma_B^{(2)} = \frac{\sigma_A^{(1)}\,\sin\left(2\,\theta_L-\theta_R\right)-\sigma_A^{(1)}\,\sin\left(\theta_R\right)+2\,\sigma_D^{(2)}\,\sin\left(\theta_R\right)}{\sin\left(2\,\theta_L-\theta_R\right)+\sin\left(\theta_R\right)},
\end{equation}
and 
\begin{equation}
\sigma_C^{(1)} = 
    -\frac{2\,\sigma_A^{(1)}\,\cos\left(\theta_L\right)-\sigma_D^{(2)}\,\cos\left(\theta_L\right)+\sigma_D^{(2)}\,\cos\left(\theta_L-2\,\theta_R\right)}{\cos\left(\theta_L-2\,\theta_R\right)+\cos\left(\theta_L\right)}.
\end{equation}
\cref{eq:temp022,eq:temp023,eq:temp024,eq:temp011} fully determine the $\sigma^{(\alpha)}$ in every region and solves the problem.

\subsection{Idealized shear band}
\label{ssec:shearband}
\begin{figure}[ht!]
\centering
\includegraphics[width=0.5\textwidth]{{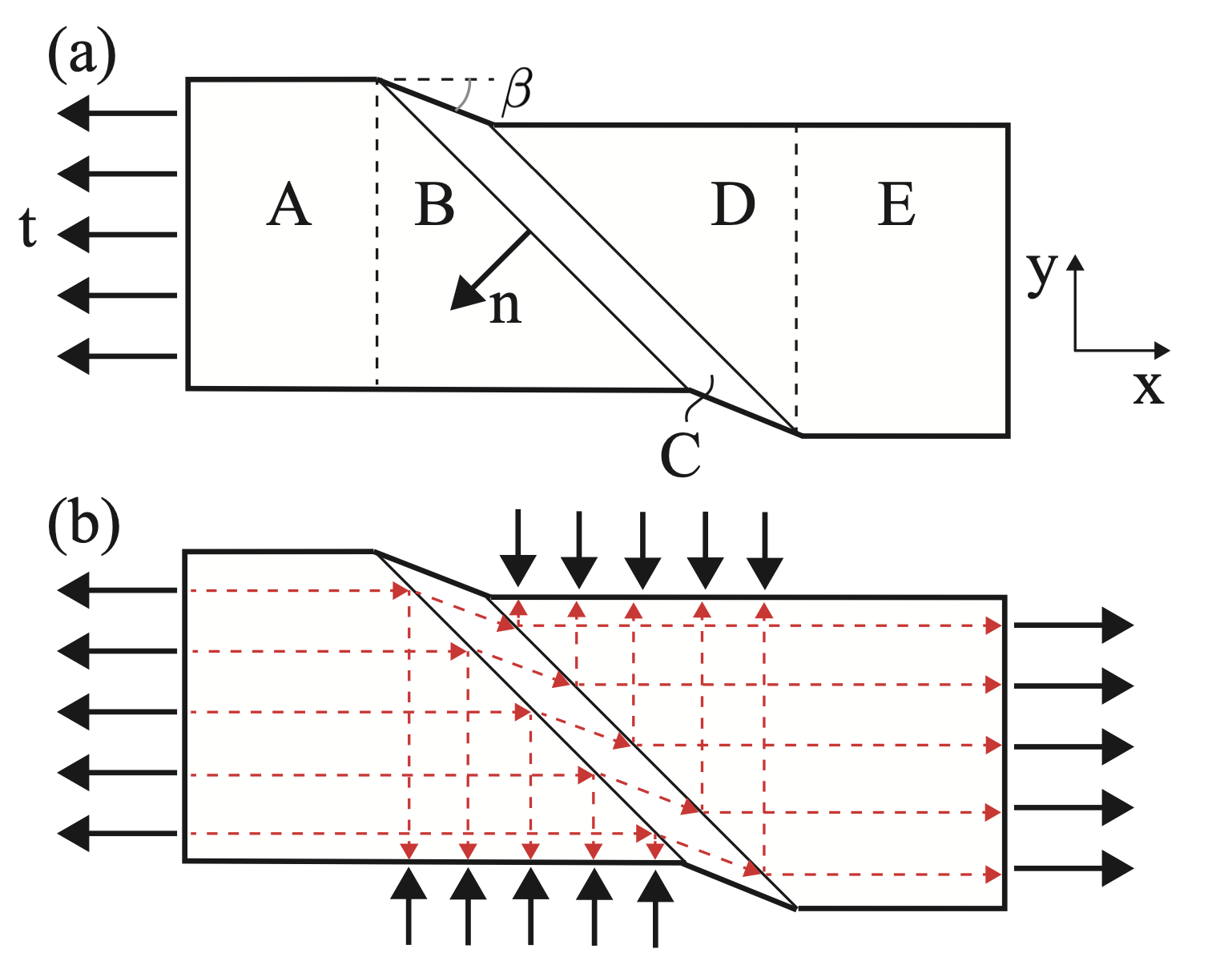}}
\caption{An idealized shear band which demonstrates the reflection and transmission of the principal stress. (a) The different regions, geometry and boundary conditions. (b) The reflection of the principal stress off the shear band, and the propagation through the shear band. The transmission coefficient gives the stresses in the shear band, and the reflection coefficient enables computation of the boundary conditions on the upper and lower side. The red lines are the characteristics. }
\label{fig:shear_band}
\end{figure}

We next consider the analytical example of an idealized shear band (\cref{fig:shear_band}a). Here, a sample is under tension with a uniform horizontal traction of magnitude $t$ on the constant $x$ boundaries (taking $x$ to be in the horizontal direction and $y$ to be in the vertical direction). The shear band is oriented at $45^o$ to the horizontal, i.e. $\bn = [-1/\sqrt{2},-1/\sqrt{2}]^\top $ (\cref{fig:shear_band}a). The shear band also has an associated angle $\beta$. We have regions A, B, D, and E outside the shear band and region C inside the shear band. We assume that the principal directions of the stress are aligned with the $x$ and $y$ axis outside the shear band so we have
\begin{equation}
\begin{split}
    \bq^{(1)}_A =\bq^{(1)}_B = \begin{bmat}
    0\\
    -1
    \end{bmat}, \qquad \bq^{(2)}_A=\bq^{(2)}_B = \begin{bmat}
    1\\
    0
    \end{bmat},\\
    \qquad \bq^{(1)}_D =\bq^{(1)}_E=\begin{bmat}
    1\\
    0
    \end{bmat}, \qquad \bq^{(2)}_D=\bq^{(2)}_E = \begin{bmat}
    0\\
    1
    \end{bmat}.
\end{split}
\end{equation}
The signs are chosen so that the stress can be computed in all regions starting from the boundary condition on the left side. We also number $\{\bq^{(i)}\}$ such that $\bq^{(2)}$ is a $90^o$ counterclockwise rotation from $\bq^{(1)}$. 

We assume that the stress state in the shear band is constant. Hence, due to the free boundary on the top and bottom of the shear band, we can deduce that
\begin{equation}
     \bq^{(1)}_C = \begin{bmat}
    -\sin(\beta)\\
    -\cos(\beta)
    \end{bmat}, \qquad \bq^{(2)}_C = \begin{bmat}
    \cos(\beta)\\
    -\sin(\beta)
    \end{bmat}.
\end{equation}
Further, we have that $\sigma_C^{(1)}=0$ because of the traction-free boundary on the top and bottom of the shear band (assuming the stress in the shear band is constant). 

Beginning at the boundary on the left side, we have
\begin{equation}
    \sigma_A^{(2)}=\sigma_B^{(2)} = t.
\end{equation}
$\sigma_A^{(1)}$ will be governed by the traction applied to the upper boundary of region A, which we assume to be zero. Hence we have
\begin{equation}
    \sigma_A^{(1)} = 0.
\end{equation}
Consider the reflection of the $\sigma_B^{(2)}$ principal stress at the B-C interface. $\bq_B^{(2)}$ and $\bq_C^{(1)}$ will be incoming to the interface with known principal stresses, and that $\bq_B^{(1)}$ and $\bq_C^{(2)}$ will be outgoing from the interface with unknown principal stresses. We solve the following equations to obtain these stresses
\begin{equation}
    \sigma^{(1)}_B\bq^{(1)}_B\bq^{(1)\top}_B\bn+\sigma^{(2)}_B\bq^{(2)}_B\bq^{(2)\top}_B\bn = \sigma^{(1)}_C\bq^{(1)}_C\bq^{(1)\top}_C\bn+\sigma^{(2)}_C\bq^{(2)}_C\bq^{(2)\top}_C\bn.
\end{equation}
This gives
\begin{equation}
    \sigma_C^{(2)} = \frac{t}{\cos^2(\beta)-\sin(\beta)\cos(\beta)}
\end{equation}
and 
\begin{equation}
    \sigma_B^{(1)} = t\frac{\sin^2(\beta)-\sin(\beta)\cos(\beta)}{\cos^2(\beta)-\sin(\beta)\cos(\beta)}.
\end{equation}
We then have the stress state in the shear band as a function $t$ and $\beta$. One can repeat the same exercise for regions D and E, or use symmetry arguments to obtain the stress states. From this we obtain
\begin{equation}
    \sigma_D^{(2)}=\sigma_B^{(1)}, \qquad \sigma_D^{(1)}=\sigma_E^{(1)} = \sigma_A^{(2)}=\sigma_B^{(2)}, \qquad \sigma_E^{(2)} = \sigma_A^{(1)} = 0.
\end{equation}
We determine that the traction applied on the bottom of region B is $-\sigma_B^{(1)}\hat{\mathbf{e}}_y$, the traction at the top of region D is $\sigma_B^{(1)}\hat{\mathbf{e}}_y$, and the traction on the right of region E is $t \hat{\mathbf{e}}_x$. One can check that this ensures the net moment applied to the sample is zero. 

\section{Finite volume formulation}
\label{sec:fv_formulation}
Here, we develop a FV discretization that (i) is able to be applied to an irregular geometry with an irregular grid; (ii) allows the direction of information propagation to be in any direction in any dimension; (iii) is stable and robust; (iv) can be applied with general boundary conditions; and (v) handles discontinuities in the principal directions of stress.

We present the FV discretization in two dimensions because this is of immediate practical importance (due to the prevalence of DIC data).  However, the generalization to three dimensions is straightforward. 

\subsection{Discretization}
\begin{figure}[ht!]
\centering
\includegraphics[width=0.5\textwidth]{{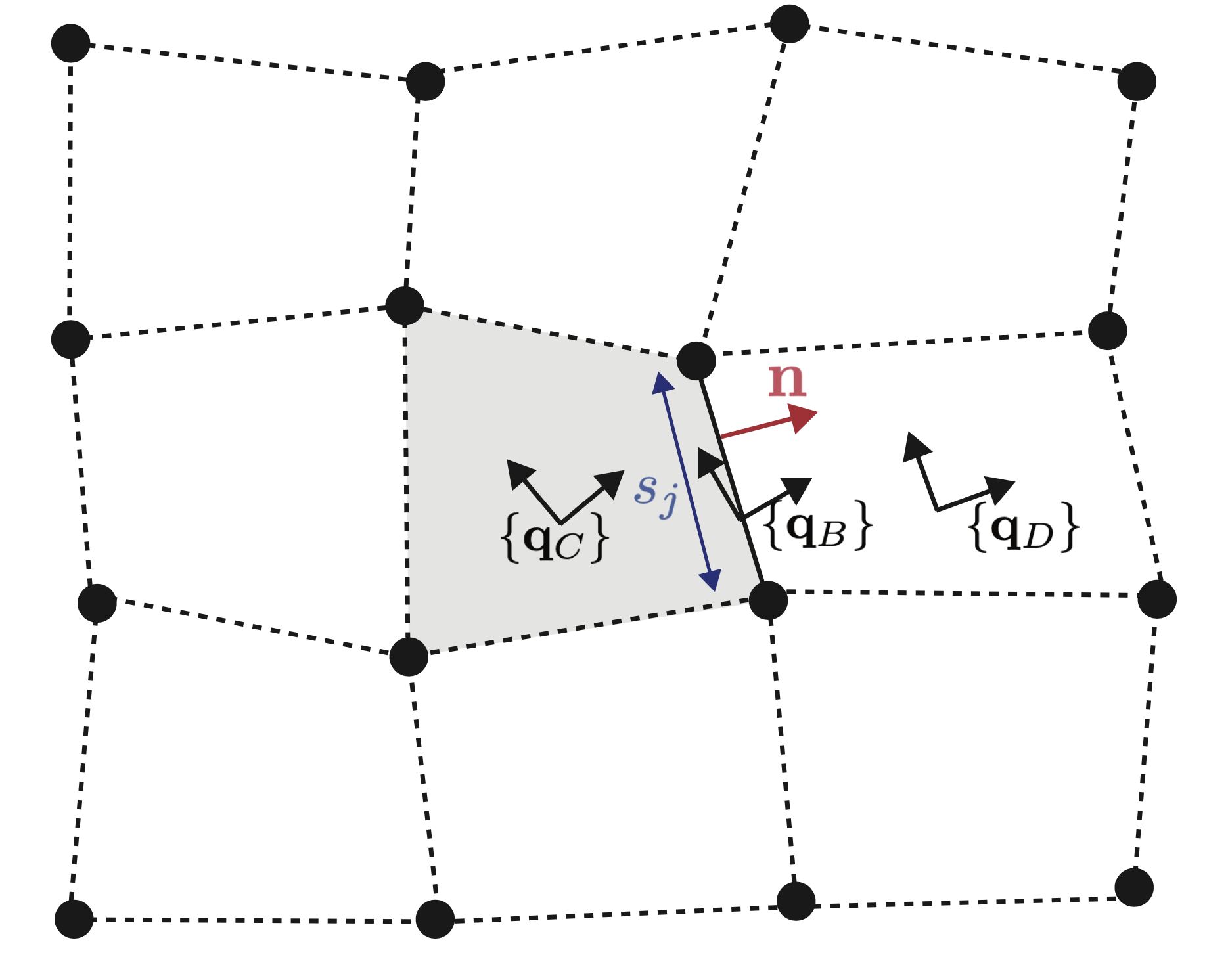}}
\caption{The FV discretization. The volumes are irregularly shaped with four nodes each. The force on each boundary must be computed.}
\label{fig:finite_volumes}
\end{figure}

First, we consider the conservation law that applies to each volume element\footnote{We use the term \textit{volume} generically, and in this case, the \textit{volume} will be two dimensional.} $\Omega$. We can simply integrate $\mathrm{div}(\bT) = 0$ over the volume then apply the divergence theorem to obtain
\begin{gequation}
\label{eq:temp012}
\int_{\partial \Omega}\bT\bn dS = 0,
\end{gequation}
where $\partial \Omega$ is the boundary and $dS$ is an infinitesimal boundary element. We specify the geometry in the deformed frame and use four sided polyhedral elements with the positions of the nodes specified at each corner (\cref{fig:finite_volumes}). We assume a constant traction $\bt^{(j)}$ on each of the four boundaries of the element. Hence, \cref{eq:temp012} becomes:
\begin{equation}
    \sum_{j,j=1}^4\bt^{(j)}s_j = 0,
    \label{eq:temp027}
\end{equation}
where $s_j$ is the length of the $j$th side of the polyhedral element. Here we have assumed a fully two-dimensional deformation where the thickness does not change, although there are straightforward modifications that can be made to account for varying thickness. 

The variable coefficient $c$ is specified to be constant inside each volume element and to be the average value of two elements on the element boundary. Note that we use the term \textit{element boundary} to refer to the boundary between two FV elements and \textit{domain boundary} to refer to the edge of the entire domain. Each element boundary will have two characteristics determined by $c$ on the boundary, denoted $\{\bq^{(\alpha)}_B\}$. Each element boundary will also have an associated normal vector $\bn$. We denote the cell in the direction of $\bn$ with a subscript $D$ and the cell in the direction of $-\bn$ with a subscript $C$ (\cref{fig:finite_volumes}). One must assign the signs of $\bq^{(\alpha)}$ such that they approximate a continuously varying orientation of each characteristic, i.e. there are no abrupt reversals in direction. One can arbitrarily choose which direction information should propagate along for each characteristic, depending on which boundary conditions one wishes to specify.

To develop a FV method, it is necessary to define a flux function on the element boundaries. Note that the flux is simply the traction on the boundary (which can be interpreted as momentum flux). Here we use a standard first order upwinding scheme with characteristic decomposition. This ensures the Courant–Friedrichs–Lewy condition is met, the consistency condition is met, and that the method is stable. $\sigma^{(\alpha)}_B$ is specified using
\begin{equation}
    \sigma^{(\alpha)}_B = \left.
  \begin{cases}
    \sigma^{(\alpha)}_C, & \text{if } \bq^{(\alpha)\top}_B\bn > 0, \\
    \sigma^{(\alpha)}_D, & \text{if } \bq^{(\alpha)\top}_B\bn < 0. \\
  \end{cases}\right.
\end{equation}
One can express the stress at the boundary utilizing the spectral decomposition. Hence, the traction along the $j$th boundary is
\begin{equation}
    \bt^{(j)} = \sigma_B^{(1)}\bq_B^{(1)}\bq_B^{(1)\top}\bn+  \sigma_B^{(2)}\bq_B^{(2)}\bq_B^{(2)\top}\bn.
    \label{eq:temp025}
\end{equation}
In the case where $\bq^{(\alpha)\top}_B\bn=0$ for some $\alpha$, $\sigma^{(\alpha)}_B$ does not need to be computed as the contribution to $\bt$ will be zero. 

The equations developed in \cref{ssec:boundary_conditions} are utilized to specify the boundary conditions. We distinguish two traction vectors at each point on the domain boundary: the traction $\bt^*$ that is given and that determines $\sigma^{(\alpha)}$ on characteristics entering the domain, and the traction $\bt^{(j)}$ that arises due to the combination of entering and leaving characteristics. For a real problem, there will be slight differences between the two tractions due to error. When both characteristics are entering the domain $\bt^{(j)}=\bt^*$. When both characteristics are leaving, the traction is determined as part of the solution and $\bt^*$ does not play a role. For one characteristic entering and one leaving, we can specify no domain boundary reflection and utilize \cref{eq:bc_norefection}. Hence we have
\begin{equation}
     \bt^{(j)} = \left.
  \begin{cases}
    \bt^* & \text{if } \bq^{(1)\top} \bn < 0 \text{  and  } \bq^{(2)\top} \bn<0,\\
    \sigma^{(1)}\bq^{(1)}\bq^{(1)\top}\bn + \sigma^{(2)}\bq^{(2)}\bq^{(2)\top}\bn & \text{if } \bq^{(1)\top} \bn > 0 \text{  and  } \bq^{(2)\top} \bn>0,\\
    \bq^{(1)}\bq^{(1)\top}\bt^*+\sigma^{(2)}\bq^{(1)}\bq^{(2)\top}\bn & \text{if } \bq^{(1)\top} \bn < 0 \text{  and  } \bq^{(2)\top} \bn>0,\\
    \bq^{(2)}\bq^{(2)\top}\bt^*+\sigma^{(1)}\bq^{(1)}\bq^{(1)\top}\bn & \text{if } \bq^{(1)\top} \bn > 0 \text{  and  } \bq^{(2)\top}\bn<0. \\
  \end{cases}\right.
  \label{eq:temp026}
\end{equation}
where $\{\bq^{(\alpha)}\}$ refer to the principal directions of the stress in the element adjacent to the domain boundary. 

For the cases considered by the authors, it was not necessary to specify the case where $\bq^{(\alpha)\top}\bn=0$ on the domain boundary because these vectors are numerically specified using floating point numbers, i.e. the dot product $\bq^{(i)\top}\bn$ will rarely be exactly zero. However, if such a case did arise one should specify that the traction component parallel to the boundary should be zero.

\cref{eq:temp025,eq:temp026} are sufficient to determine the $\bt^{(j)}$, which are then substituted into \cref{eq:temp027}. This provides two linear equations for each finite volume. Each finite volume also has two unknown principal stresses. The equations and unknowns can be assembled into a linear system of 2$n$ equations and 2$n$ unknowns, $\bA\bx=\mathbf{b}$, where $n$ is the number of finite volumes. This can then be solved using standard linear algebra algorithms. Once $\{\sigma^{(\alpha)}\}$ has been determined for each element, $\bT$ can be determined using the known principal directions. A python implementation of this code, showing how the matrix is assembled and solved, is provided in the supplementary material.

\subsection{Validation}
\label{ssec:fvvalidation}
The developed FV discretization is designed for problems with varying material properties, discontinuous strain fields, and irregular geometries. We first consider a problem developed to assess the performance of the discretization in the case of a discontinuity. Specifically, we consider an infintesimal elastic deformation with discontinuous material properties. In \cref{sssec:validation2} we briefly consider the algorithm applied to two problems from other articles to assess performance in the case of continuously varying material properties and an irregular computational mesh. The following procedure is used in each case: (i) stress and displacement/strain field is computed using a FE simulation implemented with a commercial FE solver, (ii) the kinematic field and boundary conditions are exported, (iii) the strain field is used to compute the principal directions of stress, (iv) the stress field is computed as a function of position using the developed FV discretization, (v) the stress field computed principal directions is compared to that output by the commercial FE solver and the error is quantified. In this last step we use the normalized mean absolute error (NMAE):
\begin{equation}
    \mathrm{NMAE}(\Delta\bT) = \frac{\mathrm{mean}(|\Delta \bT|)}{\mathrm{mean}(|\bT|)},
\end{equation}
where $\Delta \bT$ is the difference between the stresses in a particular element, $\bT$ is the stress computed from the FE simulation, and the mean is taken spatially over all elements in the computation.

\subsubsection{Strain discontinuity}
\label{sssec:validation1}
\begin{figure}[ht!]
\centering
\includegraphics[width=0.6\textwidth]{{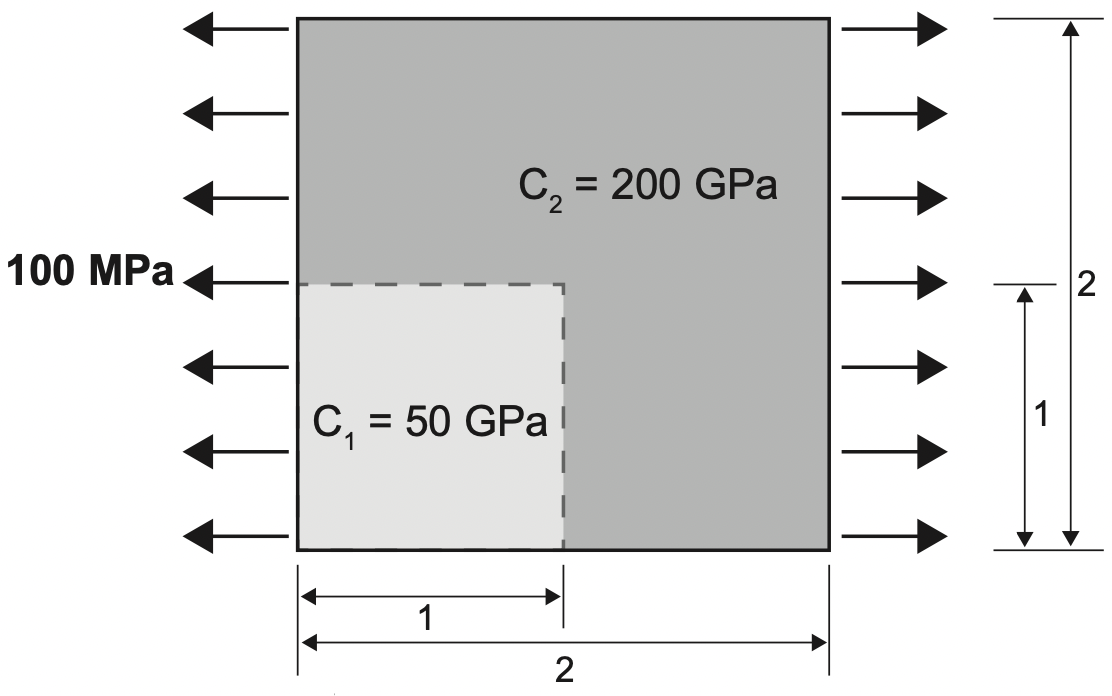}}
\caption{The problem investigated in order to validate the algorithm in the case of a discontinuity. There are two regions of the material with different elastic moduli $C$. A horizontal traction is applied on the left and right boundaries.}
\label{fig:discontinuityproblem}
\end{figure}

Consider the two-dimensional infintesimal-strain linear-elastic problem depicted in \cref{fig:discontinuityproblem}. The domain is square with two regions, one with an elastic modulus $C_1=50$~GPa, and another with an elastic modulus of $C_2=200$~GPa. The Poisson's ratio $\nu$ is $0.3$ for both regions. A constant tensile traction of 100~MPa is applied on the right and left boundaries while the upper and lower boundaries are traction free.

\begin{figure}[ht!]
\centering
\includegraphics[width=1\textwidth]{{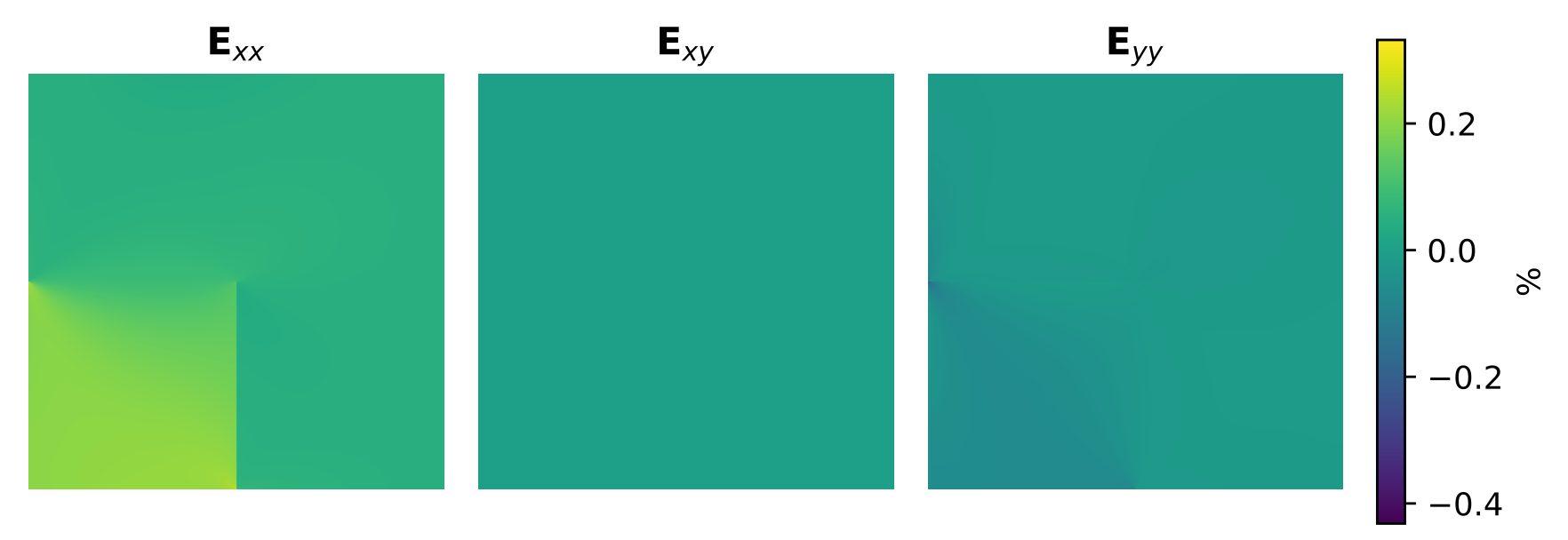}}
\caption{The strain field output from the FE simulation and input into the FV simulation.}
\label{fig:strainfield}
\end{figure}

Abaqus standard (2020) is used to conduct the FE simulations. The mesh is a regular array of square CPS4R shell elements. Simulations are conducted at a variety of resolutions, the finest being 4000$\times$4000 elements. The element boundaries coincide with the boundaries between the two domains of differing elastic moduli. \cref{fig:strainfield} shows the output strain field. 

\begin{figure}[ht!]
\centering
\includegraphics[width=0.6\textwidth]{{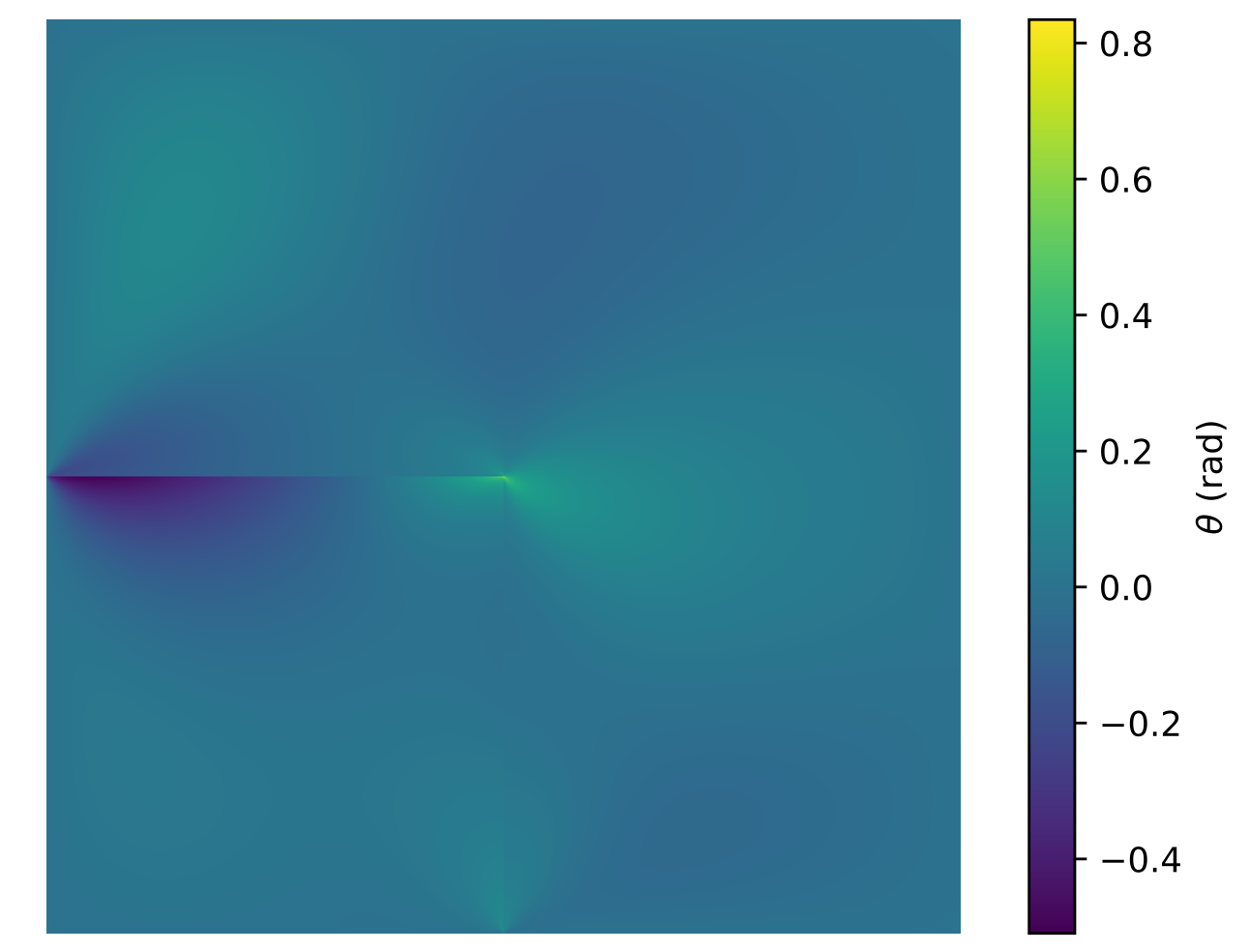}}
\caption{The orientation of the principal directions of the stress and strain as a function of position. In particular, this shows the jump between the regions with two different elastic moduli. }
\label{fig:anglediscontinuity}
\end{figure}

The principal directions are determined at each point by computing the eigenvectors of the strain output from the FE simulation. In order to assign the direction of information propagation for each volume, the following procedure is used: (i) the direction of information propagation is chosen for one "seed" volume, (ii) the direction of information propagation is chosen to be as similar as possible in adjacent cells (the dot product of vectors corresponding to the direction of information propagation is set as close as possible to 1), (iii) the prior step is repeated until the direction of information propagation is set over the entire domain. For this problem, in addition to the discontinuity present in the strain field, there is a discontinuity in the orientation of the principal directions allowing us to assess the algorithm in this case (\cref{fig:anglediscontinuity}).

The same computational mesh is used for the FV procedure as the FE procedure in order to avoid unnecessary interpolation issues. A linear system of equations is assembled and solved using a MATLAB's direct sparse matrix solver. At the finest resolution, solving the sparse system of equations took 663 s on an Intel Xeon 6226 2.70GHz CPU. Note that there were memory requirements on the order of 100 GB. Once the principal stresses are determined, the full stress tensor is computed at every point using the principal directions.
 
\begin{figure}[ht!]
\centering
\includegraphics[width=1\textwidth]{{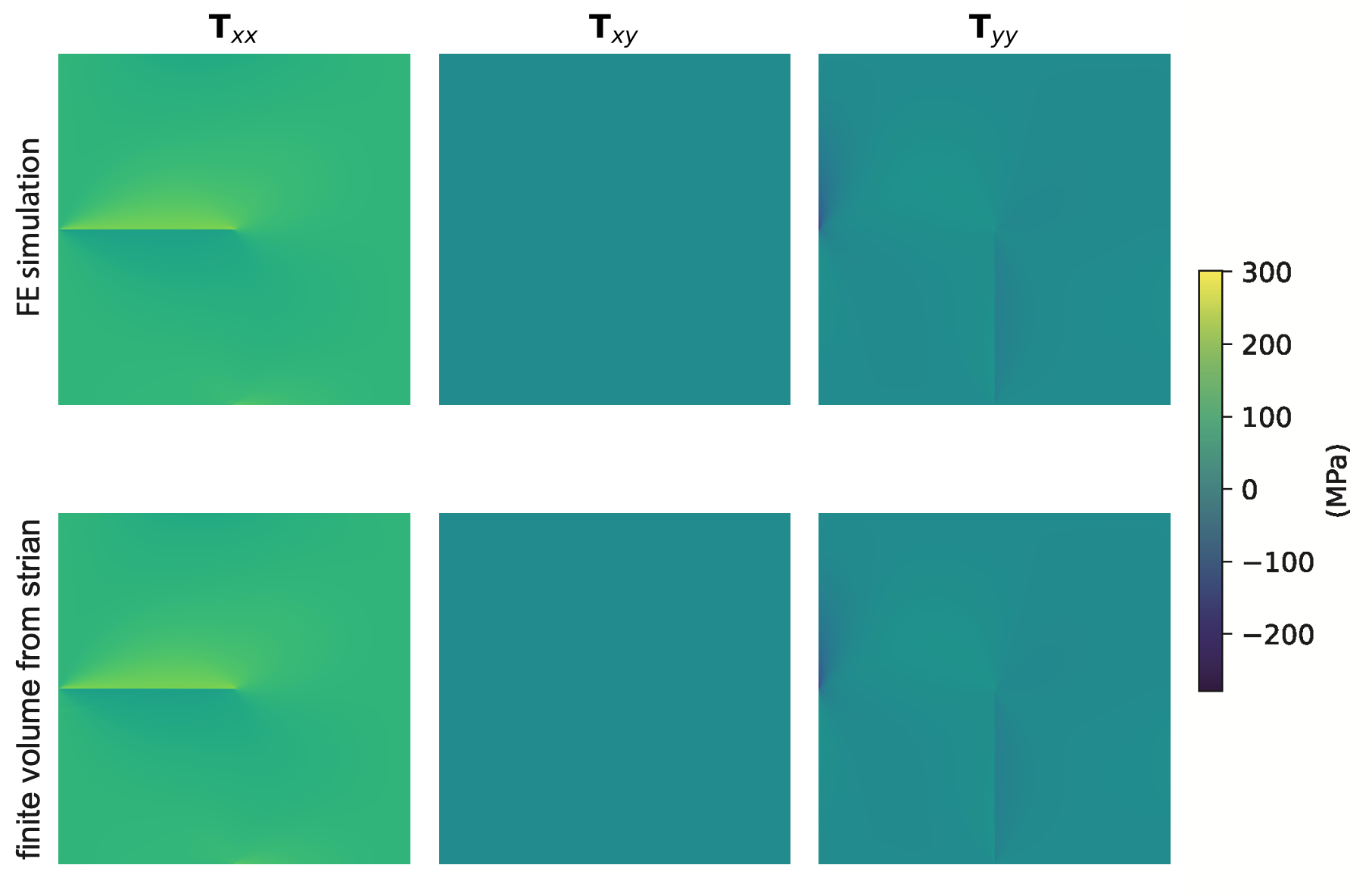}}
\caption{A comparison of the two stress fields: a field obtained via a FE simulation with specified material properties and the field obtained via FV simulation using the strain field but with no known material properties. }
\label{fig:stressfieldcompare}
\end{figure}

\begin{figure}[ht!]
\centering
\includegraphics[width=0.6\textwidth]{{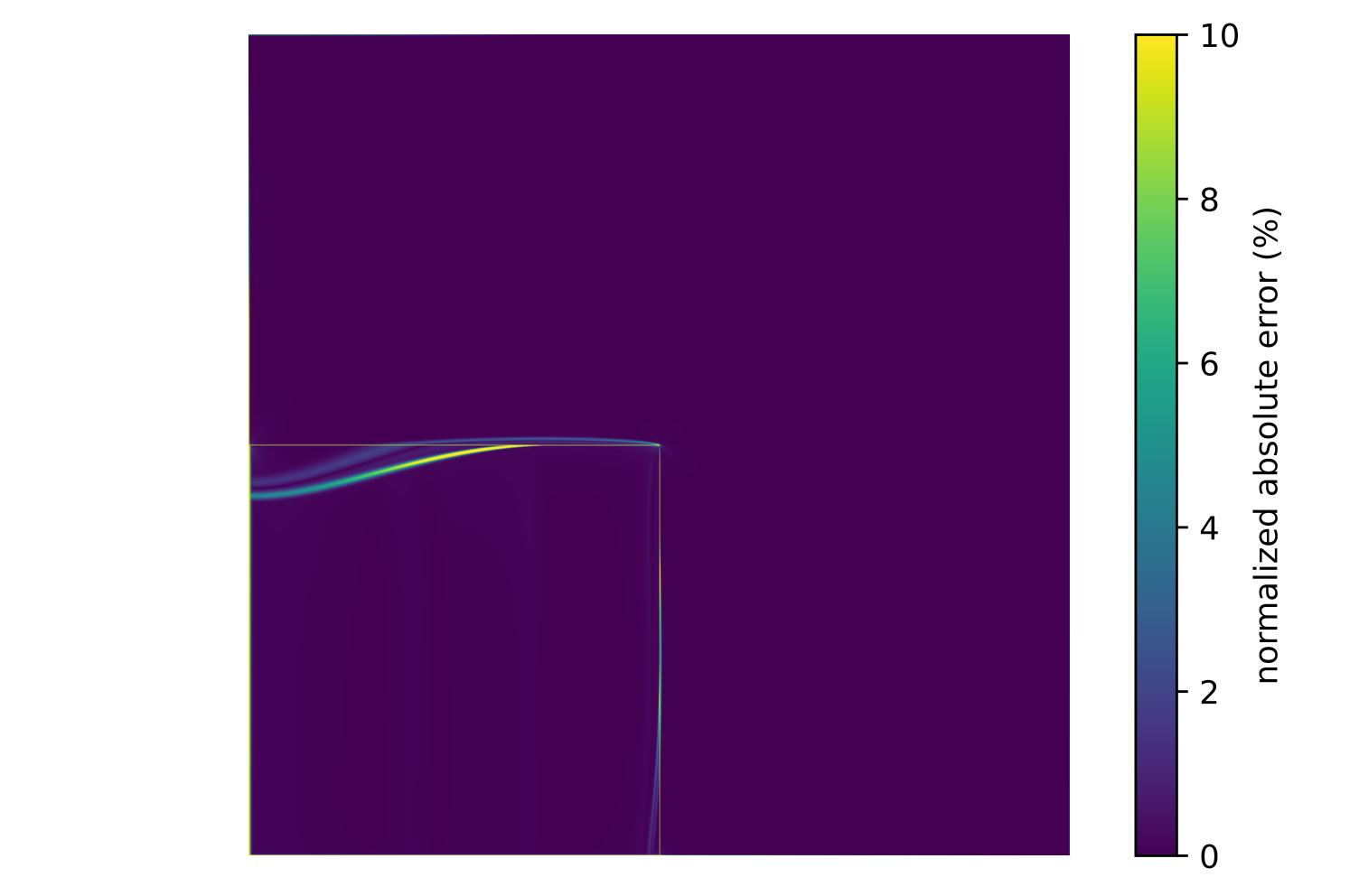}}
\caption{The normalized absolute error $|\Delta \bT|/|\bT|$ where $\Delta \bT$ is the difference between the stress computed with the FE simulation and the stress computed using the FV method and the denominator is the stress computed via the FE method at each point.}
\label{fig:NAE}
\end{figure}

\begin{figure}[ht!]
\centering
\includegraphics[width=0.6\textwidth]{{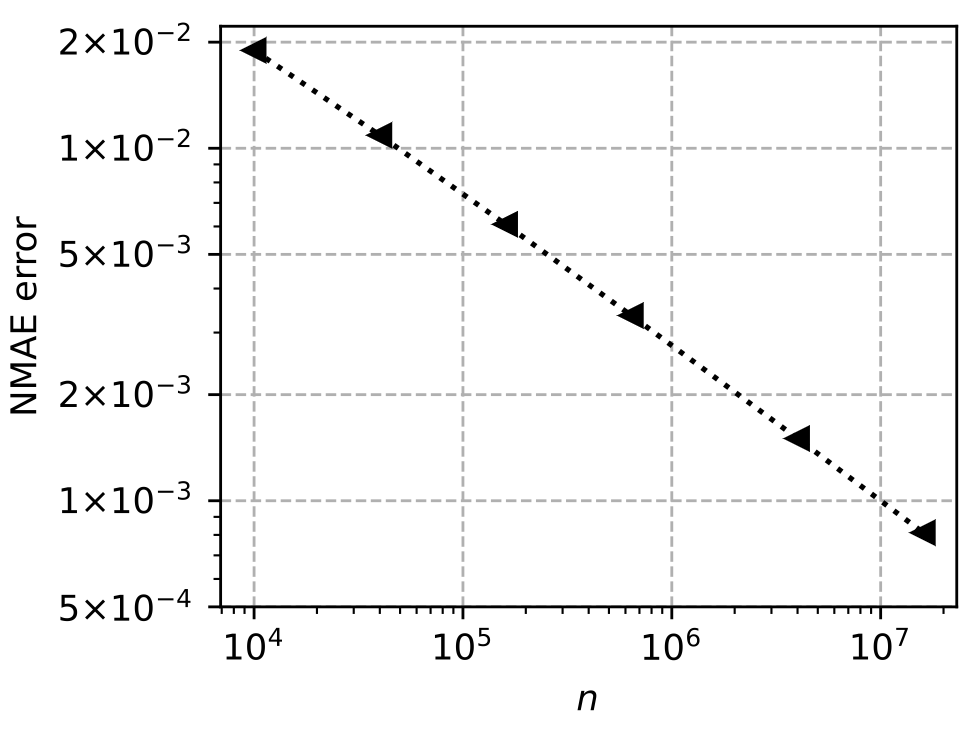}}
\caption{The normalized mean absolute error (NMAE) as a function of the number of volumes used in the FV simulation. This shows the output is converging to the true solution as the computational mesh is refined.}
\label{fig:convergence}
\end{figure}

The stress field obtained is compared to that obtained using the FE method in \cref{fig:stressfieldcompare}, with the error plotted as a function of position in \cref{fig:NAE}. There was an NMAE of 8.1$\times10^{-4}$ at the finest resolution. The solution steadily converged to the true solution as the number of volumes $n$ was increased (\cref{fig:convergence}). 

\begin{figure}[ht!]
\centering
\includegraphics[width=1\textwidth]{{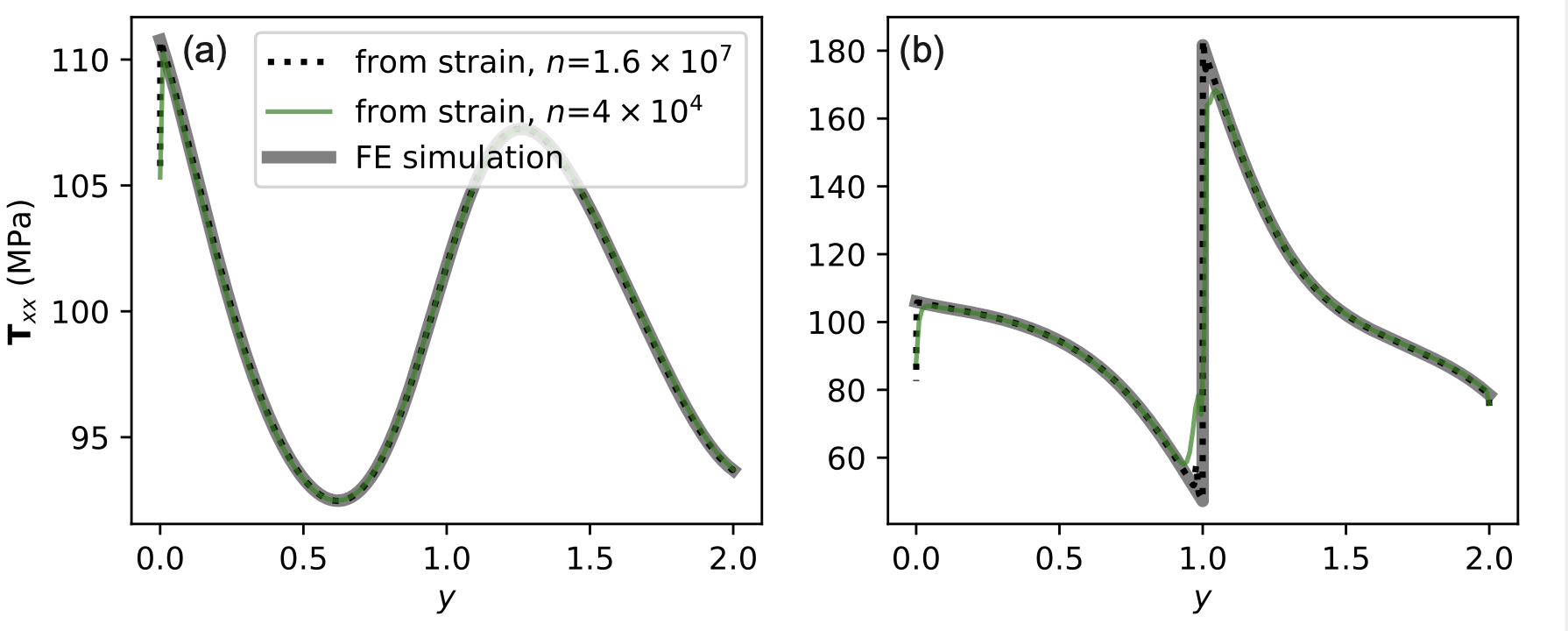}}
\caption{The stress field computed via the FE simulation compared with the FV simulation with different numbers of volumes $n$. (a) A comparison with no discontinuity in the strain with $x=1.5=3 l_x/4$. (b) A comparison with a discontinuity in the strain and principal directions ($\Delta \theta = 8.9^o$). The FE curve is taken with $n=1.6\times10^7$, however, there is no visible discrepancy between this curves and curves generated with a much smaller number of elements. }
\label{fig:linediscontinuity}
\end{figure}

The error is primarily localized at the discontinuity (\cref{fig:NAE,fig:linediscontinuity}) due to issues interpolating $\bq$ at the boundary. As the principal directions are approximately aligned with the discontinuity in some regions, small errors accumulate as the stress propagates along characteristics adjacent to the boundary. One can see some streaks in \cref{fig:NAE} which corresponds to erroneous principal stresses propagating away from the discontinuity along characteristic lines. Nevertheless, we emphasize that these errors diminish as the computational mesh is refined, and the current FV discretization appears to converge to the true solution (\cref{fig:convergence}). For example, \cref{fig:linediscontinuity} shows significant error close to the discontinuity at $n=4\times10^{4}$, but a much smaller error at $n=1.6\times 10^7$. 

One limitation of the developed FV procedure in the case of a discontinuity is that a particularly fine computational mesh is required: orders of magnitude more elements/volumes than a FE simulation for comparable accuracy. In addition, issues such as numerical diffusion may arise in other cases, as this depends on the orientation of the characteristic lines with respect to the element boundaries. The developed method is only first order accurate. This may be acceptable for many cases, however, substantial improvements to computational time, memory requirements and accuracy would be possible with a higher order discretization within the current FV framework.

\subsubsection{Other validation examples}
\label{sssec:validation2}
\begin{figure}[ht!]
\centering
\includegraphics[width=1\textwidth]{{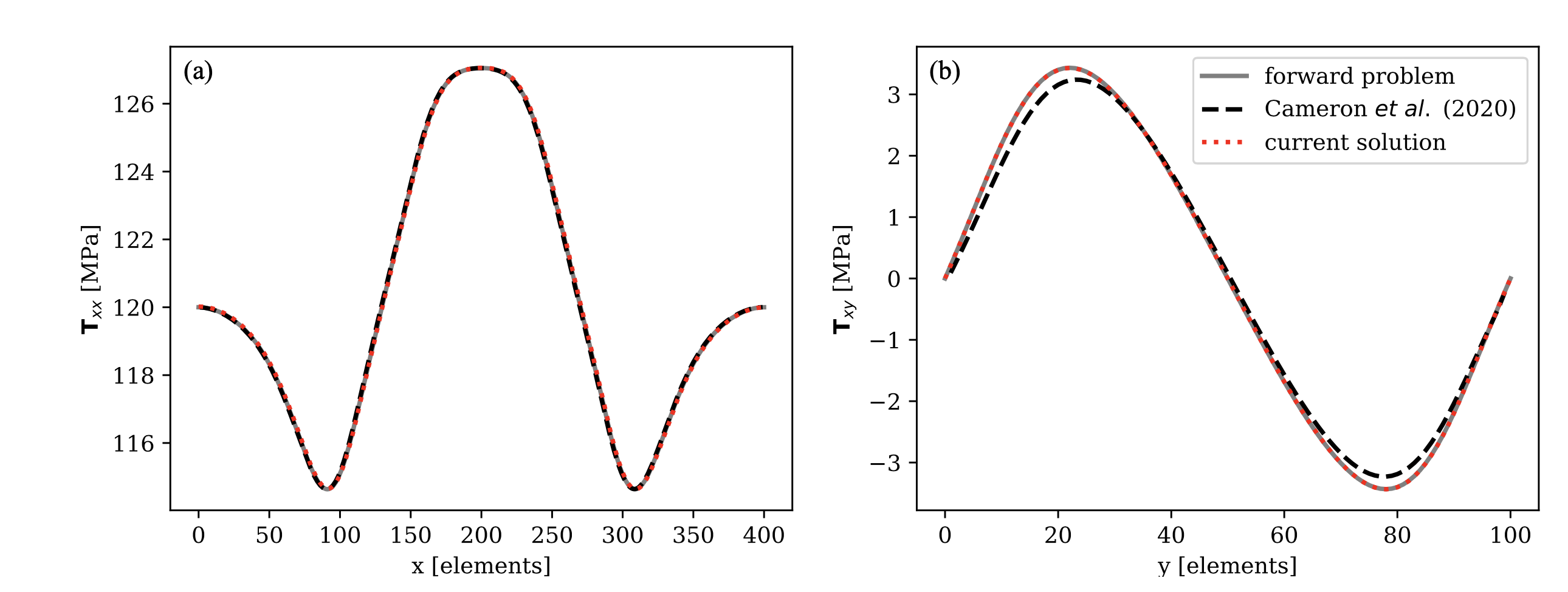}}
\caption{Comparison of the stress computed from the FE simulation described in \citet{Cameron2021} (solving the forward problem), the stress computed using the finite difference method presented in that same paper (solving the inverse problem), and the stress computed using the FV method presented in this paper (solving the inverse problem). (a) $T_{xx}$ plotted as a function of $x$ with $y=l_y/2$. (b) $T_{xy}$ plotted as a function of $y$ with $x=3l_x/4$.}\label{fig:compare_methods_line}
\end{figure}

We next consider the validation example described in detail in \citet{Cameron2021} and briefly summarized here. We have a two-dimensional rectangular sample with continuously varying material properties; specifically, yield stress and hardening coefficient. Technically, this should be interpreted as non-linear elasticity simulation because only one strain increment is simulated (see paper for details). A constant traction is applied at both ends of the sample so that the entire specimen is in tension. The simulation is conducted using 400$\times$100 square elements. A much lower spatial resolution is required compared to \cref{sssec:validation1}. For this problem, the NMAE was 9.51$\times 10^{-4}$. This was higher than the discretization presented in \citet{Cameron2021} which has a NMAE of 2.52$\times 10^{-4}$. Although that method is more accurate for this example, it is generally unstable and cannot be widely applied as frequently will violate the Courant–Friedrichs–Lewy condition. We also not that the specific error is of secondary importance compared to stability because the FV discretization can reach arbitrary accuracies through refinement of the numerical grid (see second validation example and \citet{CameronUnpublished2}). The results are shown along two line profiles in \cref{fig:compare_methods_line}.

We consider a further example obtained from \citeauthor{CameronUnpublished2} (submitted) to validate the method for an irregular geometry with an irregular mesh. This example is presented comprehensively in that article and is only briefly summarized here. There is a a two-dimensional elasto-plastic finite-strain temporal simulation of the diffuse necking in a flat tensile specimen. The simulation domain is a flat rectangular geometry with a length to width ratio of 4.5 and comprised of 7200 shell elements. The local accumulated plastic strain reaches values of 89.9~\%, and the square undeformed elements become significantly deformed, allowing the method to be tested for irregular geometries. The FV discussed in this paper was implemented at each time step, however, \bQ is determined using a more complex method discussed in \citeauthor{CameronUnpublished2} (submitted) that incorporates elasto-plasticity. The max NMAE error (computed at one increment in time but averaged over all space) is 2.1$\times 10^{-3}$, however, this is attributed to uncertainty in \bQ rather than the FV discretization because of the sensitivity of the error to the time step used.

\begin{figure}[ht!]
\centering
\label{fig:boundary_error}
\includegraphics[width=0.75\textwidth]{{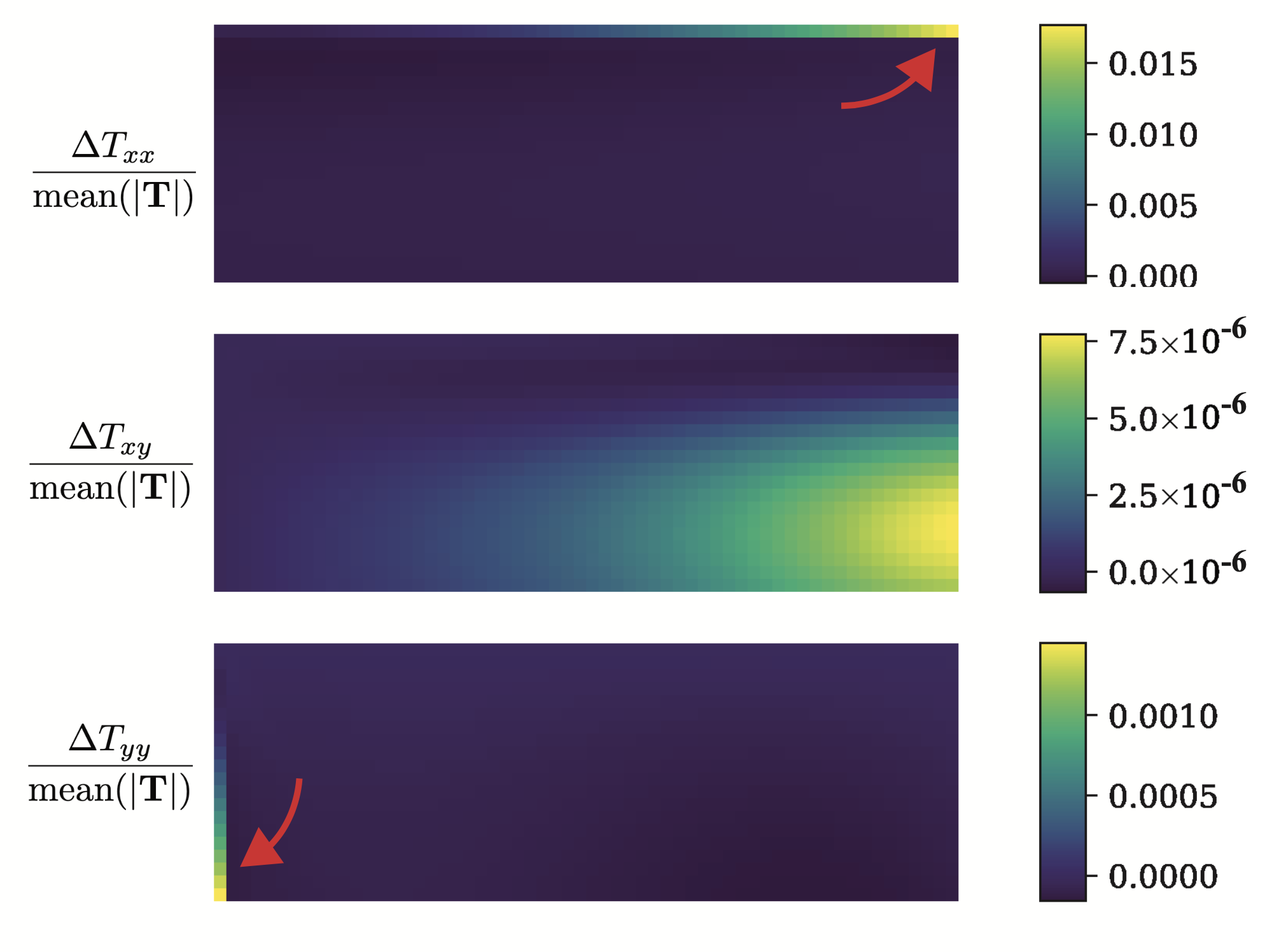}}
\caption{The error in the stress computed for the first validation example plotted in a small sub-region of the domain. This is used to show the error accumulation on the boundary. The left and upper boundaries correspond to domain boundary. The directions of information propagation are in the downward and rightward directions. }\label{fig:inverse}
\end{figure}

Note that a substantial portion of the error arises from elements directly adjacent to the boundary in all validation cases. For example, in the second validation example, the NMAE is 4.39$\times 10^{-4}$ when elements on the boundary are excluded (approximately half of the value computed when the boundaries are included). This issue remains when alternate $\bd^{*}$ vectors are used to specify the traction, or when using the finite difference discretization approach presented in \citet{Cameron2021}. Furthermore, this issue has been observed in all problems considered by the authors. The phenomenon likely arises because the characteristics at the boundary should be theoretically aligned with the boundary in many cases, whereas the characteristics of the volume element adjacent to the boundary are not. We observe that errors accumulate along the characteristic that is approximately aligned with the boundary (e.g. see \cref{fig:boundary_error}). Fortunately, from a practical perspective, one can simply ignore the information obtained close to the boundary. 

\section{Conclusion}
Analysis of the governing equations was conducted prior to developing the finite volume discretization. Key results from this analysis include:
\begin{itemize}
    \item It is mathematically shown that the characteristic directions are aligned with the principal directions of the stress. This leads to the interpretation of the propagation of principal stress components throughout the material.
    \item It is shown how a principal direction discontinuity gives rise to reflection and transmission phenomena. This is mathematically similar to the classic second order wave equation in variable wave-speed media.
    \item The correspondence between known traction at the boundary and the specific boundary conditions is elucidated for different configurations of characteristics entering and leaving the domain.
    \item The solution to the Riemann problem and an idealized shear band is given. These examples give the reader intuition for how reflection and transmission of stress apply to a physical problem.
\end{itemize}

This analysis enabled the development of a robust finite volume discretization. This is necessary for the broader applicability of the stress computation approach presented in \citet{Cameron2021}, \citet{Liu2021}, \citeauthor{CameronUnpublished2} (submitted) and the present paper. Unlike the method used in \citet{Cameron2021}, this method is stable and works for general geometries, large deformation, and more complex characteristic structures where information may propagate in any direction. Unlike the method proposed by \citet{Liu2021}, this method generalizes to any boundary conditions and is robust to discontinuities. The finite volume discretization has been validated on three examples: one infinitesimal strain problem with discontinuous material properties and strain field, one infinitesimal strain problem with material properties and strain field, and one finite strain problem with with homogeneous material
properties. There are two limitations of this analysis and discretization: (i) it does not apply to deformations with repeated stress eigenvalues, (ii) it does not apply in cases where the characteristic curves form closed loops. These limitations will likely be addressed in future work.

\appendix


\bibliographystyle{elsarticle-num-names} 
\bibliography{references}





\end{document}